\documentclass[dvipsnames,11pt]{article}
\usepackage{amsmath,amssymb,amsthm,color,bm,mathrsfs,extarrows,tikz,graphicx,mathtools,enumitem,fancybox,makecell}
\usepackage[square,sort,comma,numbers]{natbib}
\usepackage{subcaption}
\usepackage[ruled]{algorithm2e}
\usepackage[margin=1in]{geometry}
\usepackage{xcolor,bbm}
\usepackage[pagebackref]{hyperref}
\usepackage{comment}
\usepackage{soul}
\usepackage{ifthen}

\allowdisplaybreaks

\graphicspath{ {./images/} }

\setlist[itemize]{itemsep=0pt}
\setlist[enumerate]{itemsep=0pt}
\hypersetup{colorlinks=true,urlcolor=blue,linkcolor=blue,citecolor=[rgb]{.42,.56,.14},}
\usetikzlibrary{decorations.pathreplacing,decorations.markings,decorations.pathmorphing,decorations.shapes,arrows.meta,positioning,patterns}

\usepackage[capitalise,nameinlink]{cleveref}
\Crefname{lemma}{Lemma}{Lemmas}
\Crefname{fact}{Fact}{Facts}
\Crefname{theorem}{Theorem}{Theorems}
\Crefname{corollary}{Corollary}{Corollaries}
\Crefname{claim}{Claim}{Claims}
\Crefname{example}{Example}{Examples}
\Crefname{problem}{Problem}{Problems}
\Crefname{definition}{Definition}{Definitions}
\Crefname{notation}{Notation}{Notations}
\Crefname{assumption}{Assumption}{Assumptions}
\Crefname{subsection}{Subsection}{Subsections}
\Crefname{section}{Section}{Sections}
\Crefformat{equation}{(#2#1#3)}

\newtheorem{theorem}{Theorem}[section]
\newtheorem*{theorem*}{Theorem}

\newtheorem*{proposition*}{Proposition}
\newtheorem{lemma}[theorem]{Lemma}
\newtheorem*{lemma*}{Lemma}
\newtheorem{corollary}[theorem]{Corollary}
\newtheorem*{corollary*}{Corollary}
\newtheorem*{conjecture*}{Conjecture}
\newtheorem{fact}[theorem]{Fact}
\newtheorem*{fact*}{Fact}

\newtheorem*{exercise*}{Exercise}

\newtheorem*{hypothesis*}{Hypothesis}
\newtheorem{conjecture}[theorem]{Conjecture}

\theoremstyle{definition}
\newtheorem{definition}[theorem]{Definition}

\newtheorem{exercise-easy}[theorem]{Exercise}
\newtheorem{exercise-med}[theorem]{Exercise}
\newtheorem{exercise-hard}[theorem]{Exercise$^\star$}
\newtheorem{claim}[theorem]{Claim}
\newtheorem*{claim*}{Claim}

\newtheorem{remark}[theorem]{Remark}
\newtheorem*{remark*}{Remark}

\newtheorem*{observation*}{Observation}

\DeclareMathOperator*{\E}{\mathbb E}

\renewcommand{\Pr}{\operatorname*{\mathbf{Pr}}}

\newcommand{\eps}{\varepsilon}
\newcommand{\abs}[1]{\left| #1 \right|}
\newcommand{\vabs}[1]{\left\| #1 \right\|}

\newcommand{\pbra}[1]{\left( #1 \right)}
\newcommand{\sbra}[1]{\left[ #1 \right]}
\newcommand{\cbra}[1]{\left\{ #1 \right\}}
\newcommand{\floorbra}[1]{\left\lfloor #1 \right\rfloor}
\newcommand{\ceilbra}[1]{\left\lceil #1 \right\rceil}

\renewcommand{\mid}{\,\middle\vert\,}
\newcommand{\ket}[1]{\left| #1 \right\rangle}

\newcommand{\bin}{\{0,1\}}
\newcommand{\binpm}{\{\pm1\}}

\newcommand{\matsym}{\square}
\newcommand{\vecsym}{\diamondsuit}

\newcommand{\polylog}{\mathsf{polylog}}

\newcommand{\sgn}{\mathsf{sgn}}
\newcommand{\indicator}{\mathsf{1}}

\newcommand{\Phase}{\mathsf{Phase}}
\newcommand{\diag}{\mathsf{diag}}
\newcommand{\kForr}{\mathrm{forr}_k}
\newcommand{\Forr}{\mathrm{forr}}
\newcommand{\Maj}{\mathrm{Maj}}

\newcommand{\Ibm}{\bm{I}}

\newcommand{\Cbb}{\mathbb{C}}

\newcommand{\Nbb}{\mathbb{N}}
\newcommand{\Rbb}{\mathbb{R}}

\newcommand{\Acal}{\mathcal{A}}
\newcommand{\Bcal}{\mathcal{B}}
\newcommand{\Ccal}{\mathcal{C}}
\newcommand{\Dcal}{\mathcal{D}}

\newcommand{\per}{\mathrm{per}}

\newcommand{\QC}{\mathsf{Q}}

\newcommand{\sizeC}{\mathsf{Size}}
\newcommand{\subsetC}{\mathsf{Subset}}

\newcommand{\sizelC}{\mathsf{Size}_<}
\newcommand{\subsetlC}{\mathsf{Subset}_<}

\newcommand{\sizegC}{\mathsf{Size}_>}
\newcommand{\subsetgC}{\mathsf{Subset}_>}

\newboolean{includeComments}

\setboolean{includeComments}{false}

\newcommand{\kewen}[1]{
 \ifthenelse{\boolean{includeComments}}
        {{\color{red} \footnotesize Kewen: #1}}
        {}
        }
\newcommand{\mnote}[1]{
    \ifthenelse{\boolean{includeComments}}
    {{\color{blue} \footnotesize Makrand: #1}}
    {}
    }
\newcommand{\uma}[1]{
    \ifthenelse{\boolean{includeComments}}
        {{\color{brown} \footnotesize Uma: #1}}
        {}
        }
\newcommand{\avishay}[1]{
    \ifthenelse{\boolean{includeComments}}
        {{\color{purple} \footnotesize Avishay: #1}}
        {}
        }

\renewcommand{\tilde}{\widetilde}
\renewcommand{\bar}{\overline}
\renewcommand{\hat}{\widehat}

\title{The Power of Adaptivity in Quantum Query Algorithms}
\author{
Uma Girish\thanks{Princeton University. Email: \texttt{ugirish@cs.princeton.edu}.  Part of this work was done at the Simons Institute for the Theory of Computing. Supported by the Simons Collaboration on Algorithms and Geometry, a Simons Investigator Award, by the National Science Foundation grants No. CCF-1714779, CCF-2007462 and by the IBM PhD Fellowship.}
\and
Makrand Sinha\thanks{University of Illinois at Urbana-Champaign. Email: \texttt{msinha@illinois.edu}. Part of this work was done at the Simons Institute for the Theory of Computing and supported by a Simons-Berkeley Postdoctoral Fellowship.}
\and
Avishay Tal\thanks{University of California at Berkeley. Email: \texttt{atal@berkeley.edu}. Supported by a Sloan Research Fellowship and NSF CAREER Award CCF-2145474.}
\and
Kewen Wu\thanks{University of California at Berkeley. Email: \texttt{shlw\_kevin@hotmail.com}. Supported by a Sloan Research Fellowship and NSF CAREER Award CCF-2145474.}
}
\date{}

\begin{document}
\maketitle

\begin{abstract}
Motivated by limitations on the depth of near-term quantum devices, we study the depth-computation trade-off in the query model, where the depth corresponds to the number of adaptive query rounds and the computation per layer corresponds to the number of parallel queries per round. We achieve the strongest known separation between quantum algorithms with $r$ versus $r-1$ rounds of adaptivity. We do so by using the $k$-fold Forrelation problem introduced by Aaronson and Ambainis (SICOMP'18). For $k=2r$, this problem can be solved using an $r$ round quantum algorithm with only one query per round, yet we show that any $r-1$ round quantum algorithm needs an exponential (in the number of qubits) number of parallel queries per round. \medskip

Our results are proven following the Fourier analytic machinery developed in recent works on quantum-classical separations. The key new component in our result are bounds on the Fourier weights of quantum query algorithms with bounded number of rounds of adaptivity. These may be of independent interest as they distinguish the polynomials that arise from such algorithms from arbitrary bounded polynomials of the same degree.
\end{abstract}

\newpage
\section{Introduction}\label{sec:intro}

The quantum query model, also known as the \emph{black-box} or \emph{oracle} model, has been a very successful test bed to develop quantum algorithms and to give provable guarantees on speedups over classical algorithms. In this model, a quantum algorithm has ``black-box access'' to the input and is only charged for quantum queries to the input, while any intermediate computation is considered free. Most well-known quantum algorithms, such as Grover's search \cite{grover1996fast}, Deutsch-Josza's algorithm \cite{DJ92}, Bernstein-Vazirani's algorithm \cite{BV97}, Simon's Algorithm \cite{Simon97}, and Shor's period-finding algorithm \cite{Shor97}, are captured by this black-box access model. There are slightly different models of black-box access to the input and in this work, we consider the most basic access model where each query returns a \emph{bit} of the input.

\paragraph{$k$-Fold Forrelation.} Traditionally, the focus in the query model has been to compare quantum algorithms with classical ones. The culmination of this line of work led to the resolution of the following speedup question: 
\begin{center}
\it What is the largest quantum speedup that is possible over classical algorithms?
\end{center}

The motivation for this question stems from an attempt to pinpoint the exact limit of quantum speedups, and it has helped us develop a better understanding of the fundamental nature of quantum speedups. In particular, towards answering this question Aaronson and Ambainis \cite{aaronson2015forrelation} introduced the $k$-fold Forrelation problem: In this problem one evaluates a degree-$k$ polynomial that we denote by $\kForr$ (see \Cref{def:kforr}) that measures the ``Fourier correlation" between $k$ Boolean functions mapping $\{\pm1\}^m$ to $\{\pm1\}$. The algorithm can make superposition queries to any value in the truth table of these functions and 
must distinguish the case when the value of the polynomial is large from the case where it is close to zero. This defines a partial Boolean function or a promise problem. 

Letting $n=2^m$, this problem can be solved with $r = \lceil \frac{k}2 \rceil$ quantum queries with $\polylog(n)$-sized quantum circuits, while \cite{aaronson2015forrelation} also showed that it can be solved with  $O(n^{1-1/2r})$ classical queries. They conjectured that this is tight. Moreover, they also conjectured that one should be able to simulate any $r$-query quantum algorithm with $O(n^{1-1/2r})$ classical queries, making this a problem where quantum algorithms have the \emph{maximal advantage}. Up to low-order terms, the first conjecture was proven for $k$-fold Forrelation\footnote{We note that the $k=2$ case was already resolved by Aaronson and Ambainis \cite{aaronson2015forrelation} and a different proof follows from the work \cite{raz2022oracle} as well.} and its variants by Sherstov, Storozhenko, and Wu  \cite{sherstov2023optimal} and Bansal and Sinha \cite{bansal2021k}, building on the work of Raz and Tal \cite{raz2022oracle} and Tal \cite{tal2020towards}. As a complement, Bravyi, Gosset, Grier, and Schaeffer \cite{bravyi2022classical} extended the simulation result to arbitrary quantum query algorithms, showing that any $r$-query quantum algorithm can be classically simulated with $O(n^{1-1/2r})$ queries.
The $k$-fold Forrelation also turns out to be one of the most natural problems that is $\mathrm{BQP}$-complete \cite{aaronson2015forrelation} and its variants have also been proposed as candidates for other separations in quantum complexity theory \cite{KQST23}, making it a fundamental problem to study in its own right.

\paragraph{The Power of Adaptivity.} 
In this work, our focus is to identify the exact limits of quantum depth in the query model, analogous to the quantum speedup question. One of the primary motivations for studying the power of depth comes from near-term quantum hardware which is restricted to quantum circuits of small depth in order to combat decoherence due to noise. Because of depth limitations, one needs to use wider circuits with more gates in each layer to perform computation, thus making parallel operations quite desirable. This makes optimizing the depth-width trade-off a fundamental task in quantum circuit synthesis for the near-term: Reducing circuit depth allows the computation to be completed before the qubits decohere too much, but it also requires more quantum gates per layer. 

On the positive side, Cleve and Watrous \cite{cleve2000fast} showed how to implement the quantum Fourier transform in a parallel fashion, which leads to the parallelization of Shor's factoring algorithm \cite{shor1999polynomial}. Also, in a recent related work, Regev \cite{regev2023efficient} employed parallelization followed by polynomial-time classical post-processing, to  design a more efficient quantum algorithm for factoring under certain number-theoretic conjectures.
On the other hand, Moore and Nilsson \cite{moore2001parallel} conjectured that certain staircase-shaped quantum circuits cannot be efficiently parallelized. 

In the query model abstraction, the circuit depth corresponds to the number of adaptive rounds, denoted by $r$, and the circuit width corresponds to the maximal number of parallel queries, denoted by $t$, per round.
An extreme case $r=1$ is the non-adaptive quantum query algorithm, where all queries are made in parallel.
Perhaps surprisingly, van Dam \cite{van1998quantum} showed that any $n$-bit Boolean function can be computed with bounded error using only $t\le n/2+O(\sqrt n)$ non-adaptive quantum queries, which is essentially tight for total functions \cite{montanaro2010nonadaptive}.
Techniques have been developed to establish lower bounds for various problems in this non-adaptive setting \cite{nishimura2004algorithmic,koiran2010adversary,burchard2019lower}, but less is known when we have more adaptive rounds.
Zalka \cite{zalka1999grover} considered the unordered search problem on $n$-bit database and showed that $t=\Omega(n/r^2)$ is needed.
This matches the simple divide-and-search algorithm: Partition the space into $O(n/r^2)$ parts of $O(r^2)$ size each and execute Grover's algorithm \cite{grover1996fast} on each part in parallel in $r$ steps.
Jeffery, Magniez, and de Wolf \cite{jeffery2017optimal} proved tight $t=\Theta(n/r^{3/2})$ trade-off for the element distinctness problem and tight $t=\Theta(n/r^{1+1/k})$ trade-off for the $k$-sum problem.

The above results shows that being more adaptive indeed reduces the need of quantum queries. However the improvement is quite marginal: Even if we double the number of rounds, the saving is still only a constant factor.
This naturally leads to the following question:

\begin{center}
\it What is the largest possible saving in queries offered by more rounds of adaptivity?
\end{center}

\subsection{Our Results}\label{sec:our_result}

We answer the above question in the strongest sense and along the way prove structural theorems about the Fourier spectrum of polynomials that arise from low-depth quantum algorithms.

\paragraph{$r$ versus $r-1$ separation.} Our main result shows that the aforementioned $k$-fold Forrelation problem separates different levels of quantum computational power, measured in terms of adaptivity. Informally, the saving in the number of parallel queries can be unbounded, even when we just have one more adaptive round.

\begin{theorem}\label{thm:separation}
For any constant $r\ge2$, the $2r$-fold Forrelation problem on $n$-bit inputs 
\begin{enumerate}
\item\label{itm:thm:separation_1} can be solved with advantage $2^{-10r}$ by $r$ adaptive rounds of  queries with one quantum query per round, yet
\item\label{itm:thm:separation_2} any quantum query algorithm with $r-1$ adaptive rounds requires $\tilde{\Omega}(n^{1/r^2})$ parallel queries to approximate it.
\end{enumerate}
\end{theorem}

\begin{remark} ~\Cref{itm:thm:separation_2} continues to hold even in the presence of a large amount of classical pre-processing. In more detail, we consider algorithms that are allowed to first make classical queries and based on the outputs, choose a quantum algorithm to run that has $k-1$ rounds of $t$ parallel queries each. We show that any such algorithm must either make $\Omega(n^{1/(2r)})$ classical queries or $\widetilde{\Omega}(n^{1/r^2})$ quantum queries. See~\Cref{app:classical_preproc} for more details.
\end{remark}

\begin{remark}\label{rem:improvements}
We note two easy modifications of the above theorem that also follow from our work, which we do not state in the theorem statement above for brevity. First, in the first item above, one can boost the advantage of the quantum algorithm to any constant close to 1 by making $2^{O(r)}$ parallel queries per round without increasing the number of rounds since error amplification can be done by making parallel queries. Second, we can more generally obtain an $r$ versus $r'$ separation for any $r'<r$ where the lower bound in the second item improves as $r'$ decreases and is of the form $\tilde{\Omega}(n^{c(r,r')})$ where 
$$
c(r,r')=\begin{cases}
1-\frac1r & \text{ for } r'=1,\\
\frac{r-r'}{rr'+r/2}\ge\frac1{r^2} & \text{ for }2\le r'\le r-1.
\end{cases}
$$
For example, reducing the number of rounds by a factor of $2$, i.e., when $r=2r'$, gives $c(r,r') = 1/(r+1)$.
Furthermore, notice that the case when $r'=1$ corresponds to a non-adaptive lower bound: Here we obtain that any non-adaptive quantum algorithm that solves $2r$-fold Forrelation must make $\tilde{\Omega}(n^{1-1/r})$ parallel queries.
\end{remark}

\begin{remark}\label{rmk:thm:separation}
We recall that $k$-fold Forrelation is a partial function and being a partial function is necessary for \Cref{itm:thm:separation_1}.
\cite{zalka1999grover,jeffery2017optimal} showed that for any total Boolean function $f$, the number of parallel quantum queries needed with $r$ rounds is $t=\Omega(\mathsf{bs}(f)/r^2)$, where $\mathsf{bs}(f)$ is the block sensitivity complexity of $f$.
Note that Simon \cite{simon1983tight} proved that $\mathsf{bs}(f)=\Omega(\log n)$ if $f$ is a non-degenerate $n$-bit Boolean function.
This implies that $t=\Omega(\log n)$ when $r$ is a constant.
Similarly, Ambainis and de Wolf \cite{ambainis2014low} showed that any non-degenerate $n$-bit total function requires $\Omega(\log n/\log\log n)$ quantum queries in total, which implies $t=\Omega(\log n/(r\log\log n))$.
In summary, for total functions and constant rounds, the best possible separation is only logarithmic-vs-polynomial, instead of the $O(1)$-vs-polynomial separation we obtain.
\end{remark}

As mentioned previously, \Cref{itm:thm:separation_1} of \Cref{thm:separation} was already known since the work of \cite{aaronson2015forrelation} and the crux of our result is the lower bound in \Cref{itm:thm:separation_2}. Lower bounds for $k$-fold Forrelation are quite non-trivial to prove even for classical query algorithms and the known techniques rely on the polynomial method. The polynomial method cannot be directly applied since $k$-fold Forrelation is a low-degree bounded polynomial and as such one needs to find a way to distinguish it from the polynomials of much higher degree that are computed by the computational model of interest. In particular, previous works \cite{raz2022oracle, tal2020towards, bansal2021k, sherstov2023optimal} identified that if the polynomials computed by a computational model satisfy a certain refined notion of ``sparsity", in terms of \emph{bounded Fourier Growth}, then the $k$-fold Forrelation problem cannot be solved in that model.  

\paragraph{Fourier Growth of Low-depth Quantum Algorithms.} 
Recall that every Boolean function $f\colon\binpm^n\to[0,1]$ has a unique Fourier representation
$$
f(x)=\sum_{S\subseteq[n]}\hat f(S)\cdot\prod_{i\in S}x_i,
$$
where $\hat f(S)=\E[f(x)\cdot\prod_{i\in S}x_i]$ is the Fourier coefficient and the expectation is over uniform $x$ over the hypercube $\binpm^n$.
The level-$\ell$ Fourier $\ell_1$-weight $L_{1,\ell}(f)$ is defined by
$$
L_{1,\ell}(f)=\sum_{|S|=\ell}\abs{\hat f(S)}
$$
and is a measure of the capability of the function $f$ to aggregate weak signals on $\ell$ bits. 
Let $\Ccal$ be a class of Boolean functions, then the Fourier growth of $\Ccal$ refers to the scaling of $\max_{f\in\Ccal}L_{1,\ell}(f)$ when $\ell$ grows.

Following \cite{raz2022oracle,tal2020towards}, Bansal and Sinha \cite{bansal2021k} successfully related the advantage of approximating $k$-fold Forrelation for $k=2r$ with the (low-level) Fourier growth of the model of computation in question. Informally, if the Fourier weights grow slower than $(\sqrt n)^{(1-1/k)\ell}$ (which is the Fourier growth of the $k$-fold Forrelation polynomial) up to level $\ell=k^2$, then it cannot approximate $k$-fold Forrelation. As a direct application of \cite[Theorem 3.2 and Theorem 3.4]{bansal2021k}, \Cref{itm:thm:separation_2} of \Cref{thm:separation} follows from the following Fourier growth bounds.
The detailed calculations can be found in \Cref{app:thm:separation}.

\begin{theorem}\label{thm:fourier_growth}
Let $\Acal$ be a quantum query algorithm on $n$-bit inputs with arbitrarily many auxiliary qubits.
Assume $\Acal$ has $r$ adaptive rounds of $t\le n$ parallel queries.
Define $f\colon\binpm^n\to[0,1]$ by $f(x)=\Pr\sbra{\Acal\text{ accepts x}}$.
Then
$$
L_{1,\ell}(f)\le O_{r,\ell}\pbra{t^\ell\cdot \left(\sqrt{n/t}\right)^{\floorbra{ \left(1-\frac{1}{2r}\right) \ell }}}.
$$
Moreover, this bound holds when some bits of $x$ are fixed in advance.
\end{theorem}

\begin{remark}\label{rmk:tightness}
In the non-adaptive case (i.e., $r=1$), the bound in \Cref{thm:fourier_growth} can be improved (see \Cref{sec:proof_1} for detail) to
$$
L_{1,\ell}(f)\le O_\ell\pbra{t^{\ell/4}\cdot n^{\ell/4}}.
$$
This improvement implies the non-adaptive lower bound of $\tilde{\Omega}(n^{1-1/r})$ parallel queries for solving the $2r$-fold Forrelation problem, as mentioned in \Cref{rem:improvements}.
This is also tight as shown in \Cref{sec:tightness}. 
\end{remark}

The acceptance probability of any quantum query algorithm that makes $d$ queries can be expressed as a degree-$2d$ bounded polynomial. Most of the techniques in the literature do not distinguish polynomials that come from quantum algorithms from general bounded polynomials and we lack a sufficiently good understanding of such distinctions. 

Our Fourier growth bounds are far better than the bounds that can be obtained by directly applying the Fourier growth estimates for low-degree bounded polynomials \cite{DBLP:journals/eccc/IyerRRRY21,eskenazis2022learning}. Thus, this points to one way in which polynomials computed by low-depth quantum algorithms are different than general bounded polynomials of the same degree. 

\paragraph{Classically Simulating Low-depth Quantum Algorithms.} 

We mention an open problem related to the question of where the exact limits of the trade-offs between depth and the number of parallel queries lie. As mentioned before, if there is only one query per round ($t=1$), then Aaronson and Ambainis \cite{aaronson2015forrelation} conjectured that any $r$-round quantum algorithm can be simulated with $O(n^{1-1/2r})$ classical queries and this conjecture was proved by \cite{bravyi2022classical}.

Does such a classical simulation continue to exist for low-depth quantum algorithms that make multiple parallel queries per round? We believe this is the case and make the following conjecture.

\begin{conjecture}\label{conj:simulation}
Any quantum query algorithm on $n$-bit inputs with $r$ adaptive rounds and $t$ parallel queries per round can be classically simulated with $\tilde O_{t,r}\pbra{n^{1-1/2r}}$ queries.
\end{conjecture}

It is worth mentioning that the Fourier growth bounds of classical query models (aka decision trees) \cite{tal2020towards,sherstov2023optimal} scales roughly like $(D\cdot\log n)^{\ell/2}$ where $D$ is the number of classical queries.
Our Fourier bound matches the Fourier bound for decision trees of depth $\tilde O_{r,t}\pbra{n^{1-1/2r}}$ giving some support to the above conjecture.

\paragraph*{Related Works in Communication Models.}
Aside from the aforementioned results in the quantum query complexity, the round-query trade-off in the query model can also be deduced from the round-communication trade-off in the model of communication complexity.
In this model, Alice and Bob are given $n$-bit inputs $x$ and $y$ separately and their goal is to evaluate some function $F(x,y)$ by communication.

Given such a communication task $F$, we immediately get a query task $f$ by letting $z=(x,y)$ and defining $f(z)=F(x,y)$.
Then each quantum query to $z$ can be implemented in the communication setting by Alice and Bob exchanging one round of $O(\log n)$ qubits.\footnote{Here $\log n$ is required for indexing an $n$-bit string in superposition, which is not needed classically. By switching the role of Alice and Bob between communication rounds, we can simulate $r$ queries in $r$ rounds of communication and one party in the end will compute the answer.}
Therefore if $F$ requires sending $t\log n$ qubits in each round, then the corresponding $f$ requires $\Omega(t)$ parallel queries in each round.
Via this reduction, the pointer chasing problem with $r$ jumps needs $\tilde\Omega_r(n)$ parallel queries with $r-1$ adaptive rounds \cite{klauck2001interaction,jain2002quantum}, whereas it can be solved with $r$ adaptive rounds of $O(\log n)$ queries.
Since the pointer chasing problem is a total function, by \Cref{rmk:thm:separation} this logarithmic-vs-polynomial separation cannot be further improved to an $O(1)$-vs-polynomial separation.

We remark that\footnote{We thank an anonymous QIP'24 reviewer for pointing this out.} it is possible to define a variant of the pointer chasing problem which only uses one quantum query per round.
This is achieved by using the Bernstein-Vazirani trick (see \cite{Watrous09}) to encode the address of each jump by the Hadamard code.
Note that this is a partial function (due to the Bernstein-Vazirani trick), and it is conceivable that it will require $n^{\Omega(1)}$ queries if the number of adaptive rounds is reduced.
In light of this, we highlight that our results generalize to the setting of \emph{quantum query algorithms with classical preprocessing}, where the algorithm is allowed to first perform $n^{\Omega(1)}$ classical queries, then adaptively choose a quantum query algorithm with prescribed number of rounds and parallel queries. See details in \Cref{app:classical_preproc}.
In this setting, variants of the pointer chasing problem would be solved already in the classical preprocessing phase, whereas the $2r$-fold Forrelation problem still exhibits an $O(1)$-vs-polynomial separation.

\paragraph*{Related Works in Hybrid Models.}
There is another line of work on hybrid quantum-classical query algorithms that is related to the questions studied here. In particular, this line of work \cite{CoudronM20,chia2022classical,hasegawa2022optimal,AroraCCGSW23} considers the trade-off between quantum depth and the number of classical queries in a model that allows both. Although some of these works prove a fine-grained depth separation that seems similar to ours, the models considered in these works do not allow parallel queries (or only allow $\polylog(n)$-parallel queries in \cite{CoudronM20}) and they do not study the trade-offs between depth and parallel quantum queries. Consequently, these results are not comparable to ours. 

\paragraph*{Related Works in Fourier Growth.}
The study of Fourier growth dates back to Mansour \cite{mansour1995nlog} for learning theoretic purposes. More recently it has been successfully applied in the study of pseudorandomness \cite{agrawal2020coin,chattopadhyay2018pseudorandom,chattopadhyay2019pseudorandom,chattopadhyay2021fractional} and quantum-classical separations \cite{raz2022oracle,girish2022quantum,tal2020towards,bansal2021k,sherstov2023optimal,girish2023fourier}.
Moreover, Fourier growth bounds have been established for various models of computation, including Boolean circuit classes \cite{mansour1995nlog,Tal17}, branching programs \cite{RSV13,SteinkeVW17,chattopadhyay2018pseudorandom,lee2022fourier}, query models \cite{tal2020towards,bansal2021k,sherstov2023optimal,girish2021fourier}, communication models \cite{girish2022quantum,girish2023fourier}, and more.
We refer interested readers to \cite{girish2023fourier} for detailed discussion.

\paragraph*{Paper Organization.}
An overview of our Fourier growth analysis is provided in \Cref{sec:overview}.
In \Cref{sec:prelim} we define necessary notation.
\Cref{sec:fourier_growth_same} contains the full proof our results.
Missing proofs can be found in the appendix.
\section{Proof Overview}\label{sec:overview}

\subsection{High-level Proof Sketch}

\begin{figure}[ht]
\centering
\tikzset{every picture/.style={line width=0.75pt}} 

\begin{tikzpicture}[x=0.75pt,y=0.75pt,yscale=-1,xscale=1]

\draw    (212.27,139) -- (333.67,139.67) ;
\draw    (58.87,138.2) -- (189.87,138.6) ;
\draw    (211.73,145) -- (332.27,145.27) ;
\draw    (59.07,144.6) -- (190.07,145) ;
\draw    (211.87,101.27) -- (374,101.01) -- (379.5,101) ;
\draw [shift={(297.13,101.13)}, rotate = 179.91] [color={rgb, 255:red, 0; green, 0; blue, 0 }  ][line width=0.75]    (7.65,-2.3) .. controls (4.86,-0.97) and (2.31,-0.21) .. (0,0) .. controls (2.31,0.21) and (4.86,0.98) .. (7.65,2.3)   ;
\draw [shift={(380.95,101)}, rotate = 179.91] [color={rgb, 255:red, 0; green, 0; blue, 0 }  ][line width=0.75]    (7.65,-2.3) .. controls (4.86,-0.97) and (2.31,-0.21) .. (0,0) .. controls (2.31,0.21) and (4.86,0.98) .. (7.65,2.3)   ;
\draw    (59.2,100.87) -- (190.2,101.27) ;
\draw    (211.8,108.07) -- (331.67,109) ;
\draw    (59.4,107.27) -- (190.4,107.67) ;
\draw    (211.47,114.67) -- (330.67,115) ;
\draw    (59.4,113.87) -- (190.4,114.27) ;
\draw  [fill={rgb, 255:red, 255; green, 255; blue, 255 }  ,fill opacity=1 ] (129,89.8) -- (170,89.8) -- (170,149.8) -- (129,149.8) -- cycle ;
\draw  [fill={rgb, 255:red, 255; green, 255; blue, 255 }  ,fill opacity=1 ] (79.4,89.8) -- (120.4,89.8) -- (120.4,129.4) -- (79.4,129.4) -- cycle ;
\draw  [fill={rgb, 255:red, 255; green, 255; blue, 255 }  ,fill opacity=1 ] (274.6,91.4) -- (315.6,91.4) -- (315.6,151.4) -- (274.6,151.4) -- cycle ;
\draw  [fill={rgb, 255:red, 255; green, 255; blue, 255 }  ,fill opacity=1 ] (225,90.6) -- (266,90.6) -- (266,130.2) -- (225,130.2) -- cycle ;
\draw  [dash pattern={on 0.84pt off 2.51pt}]  (190.2,101.27) -- (211.87,101.27) ;
\draw  [dash pattern={on 0.84pt off 2.51pt}]  (190.4,107.67) -- (212.07,107.67) ;
\draw  [dash pattern={on 0.84pt off 2.51pt}]  (190.4,114.27) -- (212.07,114.27) ;
\draw  [dash pattern={on 0.84pt off 2.51pt}]  (189.87,138.6) -- (211.53,138.6) ;
\draw  [dash pattern={on 0.84pt off 2.51pt}]  (190.07,145) -- (211.73,145) ;
\draw  [fill={rgb, 255:red, 255; green, 255; blue, 255 }  ,fill opacity=1 ] (337.33,89) -- (363.67,89) -- (363.67,111.67) -- (337.33,111.67) -- cycle ;
\draw  [draw opacity=0] (342.72,102.24) .. controls (342.76,98.07) and (346.12,94.65) .. (350.32,94.55) .. controls (354.62,94.46) and (358.18,97.86) .. (358.28,102.16) .. controls (358.28,102.3) and (358.28,102.45) .. (358.28,102.6) -- (350.5,102.33) -- cycle ; \draw   (342.72,102.24) .. controls (342.76,98.07) and (346.12,94.65) .. (350.32,94.55) .. controls (354.62,94.46) and (358.18,97.86) .. (358.28,102.16) .. controls (358.28,102.3) and (358.28,102.45) .. (358.28,102.6) ;  
\draw    (350.5,102.33) -- (358.22,93.04) ;
\draw [shift={(359.5,91.5)}, rotate = 129.72] [color={rgb, 255:red, 0; green, 0; blue, 0 }  ][line width=0.75]    (4.37,-1.32) .. controls (2.78,-0.56) and (1.32,-0.12) .. (0,0) .. controls (1.32,0.12) and (2.78,0.56) .. (4.37,1.32)   ;

\draw (29.17,111.83) node [anchor=north west][inner sep=0.75pt]  [font=\small]  {$\ket{u}$};
\draw (86,102.33) node [anchor=north west][inner sep=0.75pt]  [font=\small]  {$O_{x}^{\otimes t} \ $};
\draw (231.33,103) node [anchor=north west][inner sep=0.75pt]  [font=\small]  {$O_{x}^{\otimes t} \ $};
\draw (138.67,105.67) node [anchor=north west][inner sep=0.75pt]  [font=\small]  {$U_{1} \ $};
\draw (283.33,106.33) node [anchor=north west][inner sep=0.75pt]  [font=\small]  {$U_{r} \ $};
\draw (384,94.5) node [anchor=north west][inner sep=0.75pt]   [align=left] {output};

\end{tikzpicture}
\caption{Quantum algorithm with $r$ adaptive rounds of $t$ parallel queries each.}
\label{figure:circuit}
\end{figure}
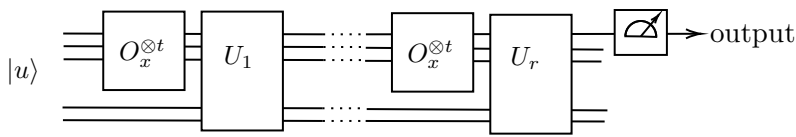

\paragraph*{Describing Quantum Algorithms with Parallel Queries.} 
Quantum algorithms which make parallel queries have the following form. First, we have an initial state $\ket{u}$; this state has some registers to index coordinates of the input and some registers for workspace. The algorithm has several rounds, where each round consists of a few parallel oracle queries followed by a unitary operator. The parallel queries are modelled by $O_x^{\otimes t}\otimes \Ibm$. Here, $O_x$ is an $(n+1)\times (n+1)$ unitary that maps $\ket{i}$ to $x_i\ket{i}$ for all $i\in[n]$ and keeps $\ket{0}$ fixed, and this is equivalent to the usual quantum query oracle. 
The operator $O_x^{\otimes t}$ implements $t$ parallel oracle queries and $\Ibm$ acts as the identity matrix on the workspace. Finally, the algorithm applies some two-outcome measurement and returns the outcome as the output. See~\Cref{figure:circuit} for depiction.

For simplicity, let us imagine that there is no workspace memory. Additionally, let us ignore the action of the oracle $O_x$ on the basis state $\ket{0}$ and treat $O_x$ as an $n\times n$ unitary matrix. These simplifications are only for the proof overview, and our proof works in full generality. In this case, the acceptance probability of this algorithm can be expressed as
\begin{equation} \label{eq:form_quantum_algorithm}
f(x)=u^\dagger O_x^{\otimes t} M_1 O_x^{\otimes t} \cdots M_{k-1} O_x^{\otimes t}  v,
\end{equation}
where $k$ is twice the number of rounds, $u=v$ corresponds to the initial state, $M_1=M_{k-1}^\dagger, M_2=M_{k-2}^\dagger,
\ldots, M_{k/2-1}=M_{ k/2+1}^\dagger$ are the $\tfrac{k}2-1$ unitary operators applied by the quantum algorithm and $M_{k/2}$ is the final measurement operator. For the rest of our proof, we can forget about the exact details of these matrices, we will only need that $M_1,\ldots,M_{k-1}$ have bounded operator norm and $u,v$ are unit vectors. 

\paragraph*{Fourier Growth of Quantum Algorithms.}
Let us now understand the Fourier growth of functions as in~\Cref{eq:form_quantum_algorithm} where $M_1,\ldots,M_{k-1}$ have bounded operator norm and $u,v$ are unit vectors.
We first set up some notation.
We use $I\in[n]^t$ to denote a $t$-tuple of elements in $[n]$. We can view $I$ as an ordered multiset of $[n]$ of size $t$ (when counted with multiplicity). Accordingly, we  use $\oplus I$ to denote the set of elements that appear an odd number of times in $I$ and use $\oplus I \oplus I'$ to denote $(\oplus I)\oplus (\oplus I')$ for $I,I'\in[n]^t$. 

When we expand the matrix multiplication in~\Cref{eq:form_quantum_algorithm}, many variables cancel out due to the identity $x_i^2=1$. 
Assume for simplicity that $u$ and $v$ are real vectors, i.e., $(u[I])^*=u[I]$.
Thus, for all $S\subseteq[n]$, the coefficient of the monomial $\prod_{i\in S}x_i$ in~\Cref{eq:form_quantum_algorithm} is given by 
\[
\widehat{f}(S)=\sum_{\substack{I_1,\ldots,I_{k}\in [n]^t\\\oplus  I_1\oplus \ldots \oplus I_{k}= S}} u[I_1] M_1[I_1,I_2]M_2[I_2,I_3]\cdots M_{k-1}[I_{k-1},I_{k}] v[I_{k}].
\]
Fix complex numbers $\alpha_S= \widehat{f}(S)^* /|\widehat{f}(S)|$ for each $S\subseteq[n]$ of size $\ell$. We wish to upper bound $L_{1,\ell}(f)=\sum_{|S|=\ell} \abs{\widehat{f}(S)}=\sum_{|S|=\ell} \alpha_S \cdot \widehat{f}(S)$, which by the above is
\begin{align}\label{eq:overview_fourier}  
L_{1,\ell}(f)&= \sum_{\substack{I_1,\ldots,I_{k}\in [n]^t\\ \abs{\oplus  I_1\oplus \cdots \oplus I_k}=\ell }} \alpha[\oplus I_1\oplus\cdots\oplus I_k]\cdot u[I_1] M_1[I_1,I_2]M_2[I_2,I_3]\cdots M_{k-1}[I_{k-1},I_{k}] v[I_{k}]. 
\end{align}

To highlight the difficulties in upper bounding~\Cref{eq:overview_fourier}, we first present a few failed approaches and then describe our high-level proof approach. First, 
let us focus on the base case $k=2$. For ease of notation, we will switch from indices $I_1,I_2$ to indices $I,J$ and from the matrix $M_1$ to $M$. Our goal is to upper bound  
\[
L_{1,\ell}(f)=\sum_{\substack{I,J\in[n]^t
\\ \abs{\oplus I \oplus J}=\ell}} \alpha[\oplus I\oplus J]\cdot u[I]M[I,J]v[J]. 
\]

One natural approach is to express $L_{1,\ell}(f)$ as a product of matrices (with bounded operator norms). One way to do this is to incorporate the phases $\alpha[\oplus I\oplus J]$ and the constraint $\abs{\oplus I \oplus J}=\ell$ into the matrix $M[I,J]$.
For instance, define $\widetilde{M}$ such that
$$
\widetilde{M}[I,J]:=\alpha[\oplus I\oplus J]\cdot \indicator[\abs{\oplus I \oplus J}=\ell]\cdot M[I,J].
$$ 
It is easy to see that $L_{1,\ell}(f)=u^\dagger \widetilde{M} v$ and consequently,  $L_{1,\ell}(f)\le \|\widetilde{M}\|$. 
What is the best upper bound that we can prove for $\|\widetilde{M}\|$? 
At first glance, it might seem that we cannot do better than $\sqrt{n^t}$. 
Indeed, given an $n^t\times n^t$ unitary matrix $M$, if we multiply each entry by arbitrary numbers in the unit disk, this could blow up the operator norm by as much as $\sqrt{n^t}$ (the Hadamard matrix gives a tight example of this). However, we can do much better. This is because the terms multiplying each entry of $M$ are highly constrained; the term multiplying the $(I,J)$-th entry depends only on $\oplus I\oplus J$. 

To get an improved bound, consider the matrix $D$ whose rows and columns are indexed by all possible $\oplus I$ and $\oplus J$ respectively, and the $(\oplus I,\oplus J)$-th entry is $\alpha[\oplus I\oplus J]\cdot \indicator\sbra{\abs{\oplus I\oplus J}=\ell}$. 
It is not too difficult to convince oneself that $\widetilde{M}$ is a sub-matrix of $M\otimes D$. Therefore, $\|\widetilde{M}\|\le \|M\|\cdot \|D\|\le \|D\|$. Now, what is the best upper bound we can show for $\|D\|$? Consider the row corresponding to $\oplus I=\emptyset$. For this row, we need to choose a column $\oplus J$ such that $\abs{\oplus J}=\ell$ and there are ${\binom n\ell}$ such columns. This already means that $\|D\|\ge \sqrt{{\binom n\ell}}$ (and this turns out to be tight). While a bound of $L_{1,\ell}(f)\le \sqrt{\binom n\ell}$ would already be a great improvement over the previous bound, it is still a trivial bound that holds for all bounded functions! Indeed, all Boolean functions which map into the complex unit disk satisfy $L_{1,\ell}(f)\le \sqrt{{\binom n\ell}}$. 

To get the optimal bound of $n^{\ell/4}\cdot t^{\ell/4}$, the idea is to reduce the operator norm of $D$. For instance, suppose we defined $\tilde{D}$ to be $D$, except that we zero out entries for which $\abs{\oplus I\setminus \oplus J}\neq \ell/2$ (or equivalently $\abs{\oplus J\setminus \oplus I}\neq \ell/2$). In this case, for any fixed $\oplus I$, the number of possibilities for $\oplus J$ is only ${\binom n{\ell/2}}\cdot {\binom t{\ell/2}}$ and we can actually prove that $\|\widetilde{D}\|\le n^{\ell/4}\cdot t^{\ell/4}$ as desired. Of course this doesn't suffice as we also need to sum over terms zeroed out. 

In the full proof, the idea is to implicitly consider all possible values of $\abs{\oplus I\setminus \oplus J}$. 
We fix any $\ell_1,\ell_2$ such that   $\abs{\oplus I\setminus \oplus J}=\ell_1$ and $\abs{\oplus J\setminus \oplus I}=\ell_2$. Since $\ell_1+\ell_2=\ell$, either (1) $\ell_1\le \ell/2$ or (2) $\ell_2\le \ell/2$. 
We will define two different matrix product decompositions to handle each of these cases separately. It will turn out that the decomposition for case (1) satisfies an operator norm bound of $n^{\ell_1/2}\cdot t^{\ell_2/2}$ and the decomposition for case (2) satisfies a bound of $n^{\ell_2/2}\cdot t^{\ell_1/2}$. Together, taking the geometric mean of the two bounds would give the desired bound of $n^{\ell/4}\cdot t^{\ell/4}$. 

We remark that our proof doesn't explicitly list out these cases; instead, it defines two different decompositions and simply takes the minimum of the two bounds which essentially captures these two cases. We describe the details of this in~\Cref{sec:proof_1}. For $k>2$, it turns out that there is a subtle but crucial over-counting issue that is too technical to describe at this point. To address this, we need to introduce new matrices in the decompositions as well as carry out a step similar to M\"obius inversion to undo the over-counting. We highlight this issue in~\Cref{sec:proof_2}.  

\subsection[Technical Proof Overview: k=2]{Technical Proof Overview: $k=2$}
\label{sec:proof_1}

Recall from~\Cref{eq:overview_fourier} that we wish to upper bound
\begin{equation} \label{eq:overview_case_1_eq1} 
L_{1,\ell}(f)=\sum_{\substack{I,J\in[n]^t
\\ \abs{\oplus I \oplus J}=\ell}} \alpha[\oplus I\oplus J]\cdot u[I]M[I,J]v[J]. 
\end{equation}  
The high-level idea is as follows. We will express $ L_{1,\ell}(f)$ as $\sum_{\substack{s_1,s_2\in \Nbb\\ s_1+s_2=\ell}}g(s)$ for some function $g(s)$, where $s=(s_1,s_2)$ and we shall group the terms based on the sizes $s_1$ and $s_2$ of the sets $\oplus I \setminus \oplus J$ and $\oplus J \setminus \oplus I$ respectively. We shall then upper bound $g(s)$ for any $s_1,s_2\in \Nbb$ satisfying $s_1+s_2=\ell$. To do this, we will express $g(s)$ in two different ways, namely, as $u^\dagger W R' v $ and as $u^\dagger W' R v $, for some matrices $W,W',R,R'$ with bounded operator norms, and we will upper bound these by $ \|u\| \|W\| \|R'\|\|v\|$ and $\|u\|\|W'\|\|R\|\|v\|$ respectively. Recall that $\|u\|=\|v\|=1$. We will show that $\|R\|,\|R'\|\le 1$ and 
\[
\|W\| \le \sqrt{{\binom n{s_2}}\cdot {\binom t{s_1}}}\quad\text{ and }\quad\|W'\| \le \sqrt{{\binom n{s_1}}\cdot {\binom t{s_2}}}.
\] 
We upper bound the minimum of the two bounds by their geometric mean and use the fact that $s_1+s_2=\ell$ to obtain 
\[ 
g(s)\le \sqrt{n^{s_2}t^{s_1}\cdot n^{s_1}t^{s_2}}=n^{\ell/4}t^{\ell/4}
\]
as desired. We now describe the function $g(s)$ and the matrices $W,W',R,R'$ in more detail. 

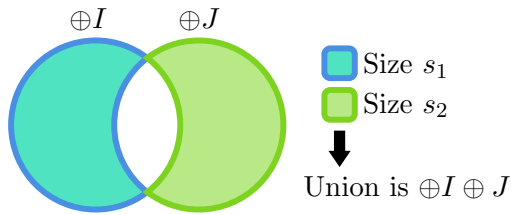
\begin{figure}[ht]
\centering
\tikzset{every picture/.style={line width=0.75pt}} 

\begin{tikzpicture}[x=0.75pt,y=0.75pt,yscale=-1,xscale=1]

\draw  [color={rgb, 255:red, 74; green, 144; blue, 226 }  ,draw opacity=1 ][fill={rgb, 255:red, 80; green, 227; blue, 194 }  ,fill opacity=1 ][line width=2.25]  (142.67,103) .. controls (152.45,103) and (161.47,106.29) .. (168.67,111.83) .. controls (158.53,119.64) and (152,131.89) .. (152,145.67) .. controls (152,159.45) and (158.53,171.7) .. (168.67,179.5) .. controls (161.47,185.04) and (152.45,188.33) .. (142.67,188.33) .. controls (119.1,188.33) and (100,169.23) .. (100,145.67) .. controls (100,122.1) and (119.1,103) .. (142.67,103) -- cycle ;
\draw  [color={rgb, 255:red, 126; green, 211; blue, 33 }  ,draw opacity=1 ][fill={rgb, 255:red, 184; green, 233; blue, 134 }  ,fill opacity=1 ][line width=2.25]  (194.67,103) .. controls (218.23,103) and (237.33,122.1) .. (237.33,145.67) .. controls (237.33,169.23) and (218.23,188.33) .. (194.67,188.33) .. controls (184.88,188.33) and (175.86,185.04) .. (168.67,179.5) .. controls (178.8,171.7) and (185.33,159.45) .. (185.33,145.67) .. controls (185.33,131.89) and (178.8,119.64) .. (168.67,111.83) .. controls (175.86,106.29) and (184.88,103) .. (194.67,103) -- cycle ;
\draw  [color={rgb, 255:red, 74; green, 144; blue, 226 }  ,draw opacity=1 ][fill={rgb, 255:red, 80; green, 227; blue, 194 }  ,fill opacity=1 ][line width=2.25]  (258,110) .. controls (258,108.34) and (259.34,107) .. (261,107) -- (270.33,107) .. controls (271.99,107) and (273.33,108.34) .. (273.33,110) -- (273.33,119) .. controls (273.33,120.66) and (271.99,122) .. (270.33,122) -- (261,122) .. controls (259.34,122) and (258,120.66) .. (258,119) -- cycle ;
\draw  [color={rgb, 255:red, 126; green, 211; blue, 33 }  ,draw opacity=1 ][fill={rgb, 255:red, 184; green, 233; blue, 134 }  ,fill opacity=1 ][line width=2.25]  (258,131) .. controls (258,129.34) and (259.34,128) .. (261,128) -- (270.33,128) .. controls (271.99,128) and (273.33,129.34) .. (273.33,131) -- (273.33,140) .. controls (273.33,141.66) and (271.99,143) .. (270.33,143) -- (261,143) .. controls (259.34,143) and (258,141.66) .. (258,140) -- cycle ;
\draw  [color={rgb, 255:red, 0; green, 0; blue, 0 }  ,draw opacity=1 ][fill={rgb, 255:red, 0; green, 0; blue, 0 }  ,fill opacity=1 ] (261,159.2) -- (263.33,159.2) -- (263.33,149) -- (268,149) -- (268,159.2) -- (270.33,159.2) -- (265.67,166) -- cycle ;

\draw (276,108) node [anchor=north west][inner sep=0.75pt]   [align=left] {Size $\displaystyle s_{1}$};
\draw (276,128) node [anchor=north west][inner sep=0.75pt]   [align=left] {Size $\displaystyle s_{2}$};
\draw (246,170) node [anchor=north west][inner sep=0.75pt]   [align=left] {Union is $\displaystyle \oplus I\oplus J$};
\draw (128,85) node [anchor=north west][inner sep=0.75pt]   [align=left] {$\displaystyle \oplus I$};
\draw (183,85) node [anchor=north west][inner sep=0.75pt]   [align=left] {$\displaystyle \oplus J$};

\end{tikzpicture}
\caption{The constraint $\sizeC^s(\oplus I,\oplus J)=1$.}
\label{figure:venn_two_sets}
\end{figure}

We group the terms in~\Cref{eq:overview_case_1_eq1} based on the sizes of $\oplus I\setminus \oplus J$ and $\oplus J\setminus \oplus I$. 
For any $(s_1, s_2)\in\Nbb\times\Nbb$, define the indicator function $\sizeC^{s}(S_1,S_2)$ for any subsets $S_1,S_2\subseteq[n]$ by
\[ 
\sizeC^{s}(S_1,S_2)=\indicator\sbra{\abs{S_1\setminus S_2}=s_1\text{ and }\abs{S_2\setminus S_1}=s_2}.
\]
We will consider $\sizeC^{s}(\oplus I,\oplus J)$ as depicted in~\Cref{figure:venn_two_sets}.

Let $g(s)$ denote the contribution to~\Cref{eq:overview_case_1_eq1} from terms satisfying $\sizeC^{s}(\oplus I,\oplus J)=1$, that is,
\[
g(s):= \sum_{I,J\in [n]^t} \sizeC^{s}(\oplus I,\oplus J)\cdot  \alpha[\oplus I \oplus J]\cdot u[I] M[I,J] v[J].
\]
From~\Cref{eq:overview_case_1_eq1}, we have $ L_{1,\ell}(f)=\sum_{\substack{s_1,s_2\in \Nbb\\ s_1+s_2=\ell}}g(s)$. 
Fix any $s_1,s_2\in \Nbb$ such that $s_1+s_2=\ell$. We will now bound $g(s)$. As described before, we will express $g(s)$ in two different ways, namely, as $u^\dagger W R' v $ and as $u^\dagger W' R v $, for some matrices $W,W',R,R'$ with bounded operator norms.

\paragraph*{Expressing $g(s)$ as $u^\dagger W R' v$.} 
The rows and columns of $W$ are indexed by $I$ and $(I',\oplus J)$ respectively, and those of $R'$ by $(I',\oplus J)$ and $J'$ respectively. These matrices are defined as follows
\begin{align*}
W[I, (I',\oplus J)]
&=\indicator\sbra{ I=  I'}\cdot \sizeC^{s}(\oplus I,\oplus J)\cdot 
\alpha[\oplus I\oplus J],\\
R'[(I',\oplus J),J']
&=\indicator\sbra{  \oplus J' =  \oplus J}\cdot  M[I',J']. 
\end{align*}
Intuitively, $W$ is a matrix that multiplies by the signs $\alpha[\oplus I\oplus J]$ as well as enforces the $\sizeC^{s}$ constraint on $\oplus I $ and $\oplus J$, and $R'$ is a matrix that implements the action of $M$, as well propagates information about $\oplus J$ backwards. This is depicted in~\Cref{figure:first_deomposition}.

\begin{figure}[ht]
\centering
\tikzset{every picture/.style={line width=0.75pt}} 

\begin{tikzpicture}[x=0.75pt,y=0.75pt,yscale=-1,xscale=1]

\draw  [line width=1.5]  (51.57,90.67) -- (129.17,90.67) -- (129.17,107.67) -- (51.57,107.67) -- cycle ;
\draw  [fill={rgb, 255:red, 248; green, 231; blue, 28 }  ,fill opacity=1 ][dash pattern={on 0.84pt off 2.51pt}] (150.67,90.67) -- (185.53,90.67) -- (185.53,110.07) -- (150.67,110.07) -- cycle ;
\draw  [color={rgb, 255:red, 0; green, 0; blue, 0 }  ,draw opacity=1 ][fill={rgb, 255:red, 248; green, 231; blue, 28 }  ,fill opacity=1 ][dash pattern={on 0.84pt off 2.51pt}] (185.53,110.07) -- (220.4,110.07) -- (220.4,129.47) -- (185.53,129.47) -- cycle ;
\draw  [fill={rgb, 255:red, 248; green, 231; blue, 28 }  ,fill opacity=1 ][dash pattern={on 0.84pt off 2.51pt}] (220.4,129.47) -- (255.27,129.47) -- (255.27,148.87) -- (220.4,148.87) -- cycle ;
\draw  [fill={rgb, 255:red, 248; green, 231; blue, 28 }  ,fill opacity=1 ][dash pattern={on 0.84pt off 2.51pt}] (255.27,148.87) -- (290.13,148.87) -- (290.13,168.27) -- (255.27,168.27) -- cycle ;
\draw  [fill={rgb, 255:red, 245; green, 166; blue, 35 }  ,fill opacity=1 ][dash pattern={on 0.84pt off 2.51pt}] (334.4,90.67) -- (334.4,125.53) -- (315,125.53) -- (315,90.67) -- cycle ;
\draw  [fill={rgb, 255:red, 245; green, 166; blue, 35 }  ,fill opacity=1 ][dash pattern={on 0.84pt off 2.51pt}] (353.8,125.53) -- (353.8,160.4) -- (334.4,160.4) -- (334.4,125.53) -- cycle ;
\draw  [fill={rgb, 255:red, 245; green, 166; blue, 35 }  ,fill opacity=1 ][dash pattern={on 0.84pt off 2.51pt}] (373.2,160.4) -- (373.2,195.27) -- (353.8,195.27) -- (353.8,160.4) -- cycle ;
\draw  [fill={rgb, 255:red, 245; green, 166; blue, 35 }  ,fill opacity=1 ][dash pattern={on 0.84pt off 2.51pt}] (392.6,195.27) -- (392.6,230.13) -- (373.2,230.13) -- (373.2,195.27) -- cycle ;
\draw  [line width=1.5]  (427.93,90.67) -- (427.93,168.33) -- (410.93,168.33) -- (410.93,90.67) -- cycle ;
\draw    (102.13,81.87) -- (127.33,82.05) ;
\draw [shift={(129.33,82.07)}, rotate = 180.42] [color={rgb, 255:red, 0; green, 0; blue, 0 }  ][line width=0.75]    (10.93,-3.29) .. controls (6.95,-1.4) and (3.31,-0.3) .. (0,0) .. controls (3.31,0.3) and (6.95,1.4) .. (10.93,3.29)   ;
\draw    (82.73,81.87) -- (53.53,81.87) ;
\draw [shift={(51.53,81.87)}, rotate = 360] [color={rgb, 255:red, 0; green, 0; blue, 0 }  ][line width=0.75]    (10.93,-3.29) .. controls (6.95,-1.4) and (3.31,-0.3) .. (0,0) .. controls (3.31,0.3) and (6.95,1.4) .. (10.93,3.29)   ;
\draw  [line width=1.5]  (290.13,90.67) -- (290.13,168.27) -- (150.67,168.27) -- (150.67,90.67) -- cycle ;
\draw  [line width=1.5]  (315,90.67) -- (392.6,90.67) -- (392.6,230.13) -- (315,230.13) -- cycle ;
\draw    (142.51,118.07) -- (142.66,92.87) ;
\draw [shift={(142.67,90.87)}, rotate = 90.32] [color={rgb, 255:red, 0; green, 0; blue, 0 }  ][line width=0.75]    (10.93,-3.29) .. controls (6.95,-1.4) and (3.31,-0.3) .. (0,0) .. controls (3.31,0.3) and (6.95,1.4) .. (10.93,3.29)   ;
\draw    (142.55,137.47) -- (142.6,166.67) ;
\draw [shift={(142.6,168.67)}, rotate = 269.9] [color={rgb, 255:red, 0; green, 0; blue, 0 }  ][line width=0.75]    (10.93,-3.29) .. controls (6.95,-1.4) and (3.31,-0.3) .. (0,0) .. controls (3.31,0.3) and (6.95,1.4) .. (10.93,3.29)   ;
\draw    (365.93,81.87) -- (391.13,82.05) ;
\draw [shift={(393.13,82.07)}, rotate = 180.42] [color={rgb, 255:red, 0; green, 0; blue, 0 }  ][line width=0.75]    (10.93,-3.29) .. controls (6.95,-1.4) and (3.31,-0.3) .. (0,0) .. controls (3.31,0.3) and (6.95,1.4) .. (10.93,3.29)   ;
\draw    (346.53,81.87) -- (317.33,81.87) ;
\draw [shift={(315.33,81.87)}, rotate = 360] [color={rgb, 255:red, 0; green, 0; blue, 0 }  ][line width=0.75]    (10.93,-3.29) .. controls (6.95,-1.4) and (3.31,-0.3) .. (0,0) .. controls (3.31,0.3) and (6.95,1.4) .. (10.93,3.29)   ;
\draw    (242.97,81.87) -- (287.27,81.87) ;
\draw [shift={(289.27,81.87)}, rotate = 180] [color={rgb, 255:red, 0; green, 0; blue, 0 }  ][line width=0.75]    (10.93,-3.29) .. controls (6.95,-1.4) and (3.31,-0.3) .. (0,0) .. controls (3.31,0.3) and (6.95,1.4) .. (10.93,3.29)   ;
\draw    (197.53,81.87) -- (153.13,81.87) ;
\draw [shift={(151.13,81.87)}, rotate = 360] [color={rgb, 255:red, 0; green, 0; blue, 0 }  ][line width=0.75]    (10.93,-3.29) .. controls (6.95,-1.4) and (3.31,-0.3) .. (0,0) .. controls (3.31,0.3) and (6.95,1.4) .. (10.93,3.29)   ;
\draw    (303.11,179.46) -- (303.26,92.87) ;
\draw [shift={(303.27,90.87)}, rotate = 90.1] [color={rgb, 255:red, 0; green, 0; blue, 0 }  ][line width=0.75]    (10.93,-3.29) .. controls (6.95,-1.4) and (3.31,-0.3) .. (0,0) .. controls (3.31,0.3) and (6.95,1.4) .. (10.93,3.29)   ;
\draw    (303.15,207.12) -- (303.2,227.87) ;
\draw [shift={(303.2,229.87)}, rotate = 269.87] [color={rgb, 255:red, 0; green, 0; blue, 0 }  ][line width=0.75]    (10.93,-3.29) .. controls (6.95,-1.4) and (3.31,-0.3) .. (0,0) .. controls (3.31,0.3) and (6.95,1.4) .. (10.93,3.29)   ;
\draw    (436.71,117.27) -- (436.86,92.07) ;
\draw [shift={(436.87,90.07)}, rotate = 90.32] [color={rgb, 255:red, 0; green, 0; blue, 0 }  ][line width=0.75]    (10.93,-3.29) .. controls (6.95,-1.4) and (3.31,-0.3) .. (0,0) .. controls (3.31,0.3) and (6.95,1.4) .. (10.93,3.29)   ;
\draw    (436.75,138.67) -- (436.8,165.87) ;
\draw [shift={(436.8,167.87)}, rotate = 269.9] [color={rgb, 255:red, 0; green, 0; blue, 0 }  ][line width=0.75]    (10.93,-3.29) .. controls (6.95,-1.4) and (3.31,-0.3) .. (0,0) .. controls (3.31,0.3) and (6.95,1.4) .. (10.93,3.29)   ;

\draw (86.6,73) node [anchor=north west][inner sep=0.75pt]   [align=left] {$\displaystyle I$};
\draw (136.2,122.2) node [anchor=north west][inner sep=0.75pt]   [align=left] {$\displaystyle I$};
\draw (352,70.4) node [anchor=north west][inner sep=0.75pt]   [align=left] {$\displaystyle J$};
\draw (271.4,185.95) node [anchor=north west][inner sep=0.75pt]   [align=left] {$\displaystyle I,\oplus J$};
\draw (431.4,120.4) node [anchor=north west][inner sep=0.75pt]   [align=left] {$\displaystyle J$};
\draw (199.4,72.35) node [anchor=north west][inner sep=0.75pt]   [align=left] {$\displaystyle I,\oplus J$};
\draw (83.8,117) node [anchor=north west][inner sep=0.75pt]   [align=left] {$\displaystyle u$};
\draw (412.6,177.4) node [anchor=north west][inner sep=0.75pt]   [align=left] {$\displaystyle v$};
\draw (209.6,180) node [anchor=north west][inner sep=0.75pt]   [align=left] {$\displaystyle W$};
\draw (344.6,239.6) node [anchor=north west][inner sep=0.75pt]   [align=left] {$\displaystyle R'$};

\end{tikzpicture}
\caption{Expressing $g(s)$ as $u^\dagger WR' v$.}
\label{figure:first_deomposition}
\end{figure}

It is not too difficult to see that indeed $g(s)= u^\dagger WR'v$. We now show the desired upper bounds of $\|R'\|\le 1$ and $\|W\|\le \sqrt{{\binom n{s_2}}\cdot {\binom t{s_1}}}.$

\begin{itemize}
    \item \textbf{Bounding $\|R'\|$:} We rearrange the columns of $R'$ according to $\oplus J$. Under this ordering of the columns, observe that $R'$ is a block diagonal matrix, where each block is a submatrix of $M$. Since $\|M\|\le 1,$ this implies that $\|R'\|\le 1$.
    \item \textbf{Bounding $\|W\|$:} We rearrange the columns of $W$ according to $I$ and with this ordering, $W$ is block-diagonal. We now use the fact that $\|W\|\le \sqrt{\|W\|_1\cdot \|W\|_\infty}$ where $\|W\|_1$ and $\|W\|_\infty$ are the max-column-norm and the max-row-norm respectively. Observe that $\|W\|_1\le 1$, since each column has at most one non-zero entry, which in turn is of unit magnitude. We now bound $\|W\|_\infty$. For any row $I\in[n]^t$, observe that there are at most ${\binom n{s_2}}\cdot {\binom t{s_1}}$ many columns $\oplus J$ such that $\sizeC^{s}(\oplus I,\oplus J)\neq 0$. Since each non-zero entry of $W$ is of unit magnitude, this implies that $\|W\|_\infty\le{\binom n{s_2}}\cdot {\binom t{s_1}}$.
    This gives us the desired bound of
\begin{equation}\label{eq:overview_case_1_bound_1} \|W\|\le\sqrt{{\binom n{s_2}}\cdot {\binom t{s_1}}}. 
\end{equation}
\end{itemize} 

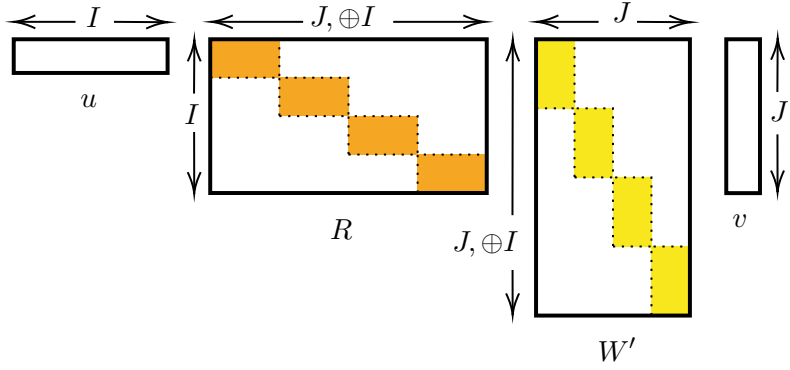
\begin{figure}[ht]
\centering
\tikzset{every picture/.style={line width=0.75pt}} 

\begin{tikzpicture}[x=0.75pt,y=0.75pt,yscale=-1,xscale=1]

\draw  [line width=1.5]  (51.57,90.67) -- (129.17,90.67) -- (129.17,107.67) -- (51.57,107.67) -- cycle ;
\draw  [fill={rgb, 255:red, 245; green, 166; blue, 35 }  ,fill opacity=1 ][dash pattern={on 0.84pt off 2.51pt}] (150.67,90.67) -- (185.53,90.67) -- (185.53,110.07) -- (150.67,110.07) -- cycle ;
\draw  [color={rgb, 255:red, 0; green, 0; blue, 0 }  ,draw opacity=1 ][fill={rgb, 255:red, 245; green, 166; blue, 35 }  ,fill opacity=1 ][dash pattern={on 0.84pt off 2.51pt}] (185.53,110.07) -- (220.4,110.07) -- (220.4,129.47) -- (185.53,129.47) -- cycle ;
\draw  [fill={rgb, 255:red, 245; green, 166; blue, 35 }  ,fill opacity=1 ][dash pattern={on 0.84pt off 2.51pt}] (220.4,129.47) -- (255.27,129.47) -- (255.27,148.87) -- (220.4,148.87) -- cycle ;
\draw  [fill={rgb, 255:red, 245; green, 166; blue, 35 }  ,fill opacity=1 ][dash pattern={on 0.84pt off 2.51pt}] (255.27,148.87) -- (290.13,148.87) -- (290.13,168.27) -- (255.27,168.27) -- cycle ;
\draw  [fill={rgb, 255:red, 248; green, 231; blue, 28 }  ,fill opacity=1 ][dash pattern={on 0.84pt off 2.51pt}] (334.4,90.67) -- (334.4,125.53) -- (315,125.53) -- (315,90.67) -- cycle ;
\draw  [fill={rgb, 255:red, 248; green, 231; blue, 28 }  ,fill opacity=1 ][dash pattern={on 0.84pt off 2.51pt}] (353.8,125.53) -- (353.8,160.4) -- (334.4,160.4) -- (334.4,125.53) -- cycle ;
\draw  [fill={rgb, 255:red, 248; green, 231; blue, 28 }  ,fill opacity=1 ][dash pattern={on 0.84pt off 2.51pt}] (373.2,160.4) -- (373.2,195.27) -- (353.8,195.27) -- (353.8,160.4) -- cycle ;
\draw  [fill={rgb, 255:red, 248; green, 231; blue, 28 }  ,fill opacity=1 ][dash pattern={on 0.84pt off 2.51pt}] (392.6,195.27) -- (392.6,230.13) -- (373.2,230.13) -- (373.2,195.27) -- cycle ;
\draw  [line width=1.5]  (427.93,90.67) -- (427.93,168.33) -- (410.93,168.33) -- (410.93,90.67) -- cycle ;
\draw    (102.13,81.87) -- (127.33,82.05) ;
\draw [shift={(129.33,82.07)}, rotate = 180.42] [color={rgb, 255:red, 0; green, 0; blue, 0 }  ][line width=0.75]    (10.93,-3.29) .. controls (6.95,-1.4) and (3.31,-0.3) .. (0,0) .. controls (3.31,0.3) and (6.95,1.4) .. (10.93,3.29)   ;
\draw    (82.73,81.87) -- (53.53,81.87) ;
\draw [shift={(51.53,81.87)}, rotate = 360] [color={rgb, 255:red, 0; green, 0; blue, 0 }  ][line width=0.75]    (10.93,-3.29) .. controls (6.95,-1.4) and (3.31,-0.3) .. (0,0) .. controls (3.31,0.3) and (6.95,1.4) .. (10.93,3.29)   ;
\draw  [line width=1.5]  (290.13,90.67) -- (290.13,168.27) -- (150.67,168.27) -- (150.67,90.67) -- cycle ;
\draw  [line width=1.5]  (315,90.67) -- (392.6,90.67) -- (392.6,230.13) -- (315,230.13) -- cycle ;
\draw    (142.51,118.07) -- (142.66,92.87) ;
\draw [shift={(142.67,90.87)}, rotate = 90.32] [color={rgb, 255:red, 0; green, 0; blue, 0 }  ][line width=0.75]    (10.93,-3.29) .. controls (6.95,-1.4) and (3.31,-0.3) .. (0,0) .. controls (3.31,0.3) and (6.95,1.4) .. (10.93,3.29)   ;
\draw    (142.55,137.47) -- (142.6,166.67) ;
\draw [shift={(142.6,168.67)}, rotate = 269.9] [color={rgb, 255:red, 0; green, 0; blue, 0 }  ][line width=0.75]    (10.93,-3.29) .. controls (6.95,-1.4) and (3.31,-0.3) .. (0,0) .. controls (3.31,0.3) and (6.95,1.4) .. (10.93,3.29)   ;
\draw    (365.93,81.87) -- (391.13,82.05) ;
\draw [shift={(393.13,82.07)}, rotate = 180.42] [color={rgb, 255:red, 0; green, 0; blue, 0 }  ][line width=0.75]    (10.93,-3.29) .. controls (6.95,-1.4) and (3.31,-0.3) .. (0,0) .. controls (3.31,0.3) and (6.95,1.4) .. (10.93,3.29)   ;
\draw    (346.53,81.87) -- (317.33,81.87) ;
\draw [shift={(315.33,81.87)}, rotate = 360] [color={rgb, 255:red, 0; green, 0; blue, 0 }  ][line width=0.75]    (10.93,-3.29) .. controls (6.95,-1.4) and (3.31,-0.3) .. (0,0) .. controls (3.31,0.3) and (6.95,1.4) .. (10.93,3.29)   ;
\draw    (242.97,81.87) -- (287.27,81.87) ;
\draw [shift={(289.27,81.87)}, rotate = 180] [color={rgb, 255:red, 0; green, 0; blue, 0 }  ][line width=0.75]    (10.93,-3.29) .. controls (6.95,-1.4) and (3.31,-0.3) .. (0,0) .. controls (3.31,0.3) and (6.95,1.4) .. (10.93,3.29)   ;
\draw    (197.53,81.87) -- (153.13,81.87) ;
\draw [shift={(151.13,81.87)}, rotate = 360] [color={rgb, 255:red, 0; green, 0; blue, 0 }  ][line width=0.75]    (10.93,-3.29) .. controls (6.95,-1.4) and (3.31,-0.3) .. (0,0) .. controls (3.31,0.3) and (6.95,1.4) .. (10.93,3.29)   ;
\draw    (303.11,179.46) -- (303.26,92.87) ;
\draw [shift={(303.27,90.87)}, rotate = 90.1] [color={rgb, 255:red, 0; green, 0; blue, 0 }  ][line width=0.75]    (10.93,-3.29) .. controls (6.95,-1.4) and (3.31,-0.3) .. (0,0) .. controls (3.31,0.3) and (6.95,1.4) .. (10.93,3.29)   ;
\draw    (303.15,207.12) -- (303.2,227.87) ;
\draw [shift={(303.2,229.87)}, rotate = 269.87] [color={rgb, 255:red, 0; green, 0; blue, 0 }  ][line width=0.75]    (10.93,-3.29) .. controls (6.95,-1.4) and (3.31,-0.3) .. (0,0) .. controls (3.31,0.3) and (6.95,1.4) .. (10.93,3.29)   ;
\draw    (436.71,117.27) -- (436.86,92.07) ;
\draw [shift={(436.87,90.07)}, rotate = 90.32] [color={rgb, 255:red, 0; green, 0; blue, 0 }  ][line width=0.75]    (10.93,-3.29) .. controls (6.95,-1.4) and (3.31,-0.3) .. (0,0) .. controls (3.31,0.3) and (6.95,1.4) .. (10.93,3.29)   ;
\draw    (436.75,138.67) -- (436.8,165.87) ;
\draw [shift={(436.8,167.87)}, rotate = 269.9] [color={rgb, 255:red, 0; green, 0; blue, 0 }  ][line width=0.75]    (10.93,-3.29) .. controls (6.95,-1.4) and (3.31,-0.3) .. (0,0) .. controls (3.31,0.3) and (6.95,1.4) .. (10.93,3.29)   ;

\draw (86.6,73) node [anchor=north west][inner sep=0.75pt]   [align=left] {$\displaystyle I$};
\draw (136.2,122.2) node [anchor=north west][inner sep=0.75pt]   [align=left] {$\displaystyle I$};
\draw (352,70.4) node [anchor=north west][inner sep=0.75pt]   [align=left] {$\displaystyle J$};
\draw (271.4,185.95) node [anchor=north west][inner sep=0.75pt]   [align=left] {$\displaystyle J,\oplus I$};
\draw (431.4,120.4) node [anchor=north west][inner sep=0.75pt]   [align=left] {$\displaystyle J$};
\draw (199.4,72.35) node [anchor=north west][inner sep=0.75pt]   [align=left] {$\displaystyle J,\oplus I$};
\draw (83.8,117) node [anchor=north west][inner sep=0.75pt]   [align=left] {$\displaystyle u$};
\draw (412.6,177.4) node [anchor=north west][inner sep=0.75pt]   [align=left] {$\displaystyle v$};
\draw (209.6,180) node [anchor=north west][inner sep=0.75pt]   [align=left] {$\displaystyle R$};
\draw (344.6,239.6) node [anchor=north west][inner sep=0.75pt]   [align=left] {$\displaystyle W'$};

\end{tikzpicture}
\caption{Expressing $g(s)$ as $u^\dagger R W' v$.}
\label{figure:second_decomposition}
\end{figure}

\paragraph*{Expressing $g(s)$ as $u^\dagger  RW' v$.} 
The rows and columns of $R$ are indexed by $I$ and $(J',\oplus I')$ respectively   and those of $W'$ are indexed by $(J',\oplus I')$ and $J$ respectively, and
\begin{align*} 
W'[(J',\oplus I'),J]
&=\indicator\sbra{ J=  J'}\cdot \sizeC^{s}(\oplus I',\oplus J) \cdot 
\alpha[\oplus I'\oplus J],\\
R[I,(J',\oplus I')]
&=\indicator\sbra{  \oplus I =  \oplus I'}\cdot  M[I,J'].
\end{align*}
Here, $W'$ implements $\alpha[\oplus I\oplus J]$ as well as enforces the $\sizeC^{s}$ constraint on $\oplus I $ and $\oplus J$, and $R$ implements the action of $M$, as well propagates information about $\oplus I$ forward. This is depicted in~\Cref{figure:second_decomposition}. 
A calculation similar to the previous case implies the desired bound of 
\begin{equation}\label{eq:overview_case_1_bound_2}
\|W'\|\le\sqrt{{\binom n{s_1}}\cdot {\binom t{s_2}}}.
\end{equation} 
This completes the proof overview for $k=2$.

\subsection[Technical Proof Overview: k=3]{Technical Proof Overview: $k=3$}\label{sec:proof_2} 
For simplicity of notation, we will switch from indices $I_1,I_2,I_3$ to indices $I,J,K$. We need to upper bound
\[ 
L_{1,\ell}(f)=\sum_{\substack{I,J,K\in[n]^t\\ \abs{\oplus I\oplus J\oplus K}=\ell }} \alpha[\oplus I\oplus J \oplus K]\cdot u[I] M_1[I,J] M_2[J,K] v[K]. 
\]
As before, we will express $L_{1,\ell}(f)$ as $\sum_{\substack{s_1,\ldots,s_4\in \Nbb\\ s_1+\cdots+s_4=\ell}}g(s)$ grouping terms based on sizes of certain sets and in order to bound each $g(s)$, we will try to express it in three different ways  as $u^\dagger W_1 R'_1 R'_2 v$, $u^\dagger R_1 W_2 R'_2 v$ and $u^\dagger R_1 R_2 W_3 v$. It will turn out that $\|R_1\|,\|R'_1\|,\|R_2\|,\|R'_2\|\le 1$ and that 
\begin{equation}\label{eq:intro_bound_want} 
\|W_1\|\le (n/t)^{(s_2+s_3)/2}\cdot t^{\ell},\quad\quad \|W_2\|\le (n/t)^{(s_1+s_3)/2}\cdot t^{\ell},\quad \quad \|W_3\|\le (n/t)^{(s_1+s_2)/2}\cdot t^{\ell}.
\end{equation}
Since $s_1+s_2+s_3\le \ell$, taking the minimum of the three bounds would give us the desired bound of $(n/t)^{\ell/3}\cdot t^{\ell}$. There is an issue that comes up that we will later highlight. To describe it now in a nutshell, it turns out we \emph{cannot} express $g(s)$ in the form of a matrix product with  operator norms bounded as desired. Nevertheless, with some additional work, we can express a different function $h(s)$ in this form, furthermore, $h(s)=\sum_{s'}P[s,s']g(s')$ for some invertible matrix $P$ such that $P^{-1}$ has bounded norms. Therefore, using bounds on $h(s)$, we can derive the desired bounds on $g(s)$. We describe all this in more detail.

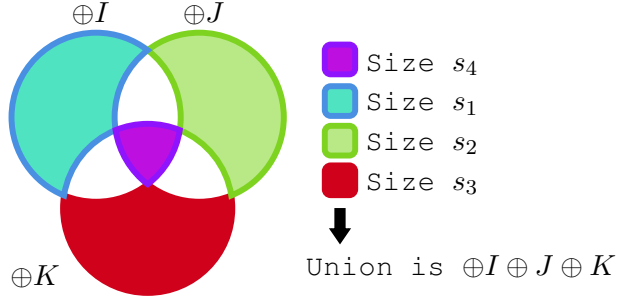
\begin{figure}[ht]
\centering
\tikzset{every picture/.style={line width=0.75pt}} 

\begin{tikzpicture}[x=0.75pt,y=0.75pt,yscale=-1,xscale=1]

\draw  [color={rgb, 255:red, 208; green, 2; blue, 27 }  ,draw opacity=1 ][fill={rgb, 255:red, 208; green, 2; blue, 27 }  ,fill opacity=0.63 ][line width=2.25]  (211.33,191.67) .. controls (211.33,215.23) and (192.23,234.33) .. (168.67,234.33) .. controls (145.1,234.33) and (126,215.23) .. (126,191.67) .. controls (126,189.45) and (126.17,187.28) .. (126.49,185.16) .. controls (131.48,187.21) and (136.94,188.33) .. (142.67,188.33) .. controls (152.45,188.33) and (161.47,185.04) .. (168.67,179.5) .. controls (175.86,185.04) and (184.88,188.33) .. (194.67,188.33) .. controls (200.39,188.33) and (205.85,187.21) .. (210.84,185.16) .. controls (211.17,187.28) and (211.33,189.45) .. (211.33,191.67) -- cycle ;
\draw  [color={rgb, 255:red, 126; green, 211; blue, 33 }  ,draw opacity=1 ][fill={rgb, 255:red, 184; green, 233; blue, 134 }  ,fill opacity=1 ][line width=2.25]  (194.67,103) .. controls (218.23,103) and (237.33,122.1) .. (237.33,145.67) .. controls (237.33,163.51) and (226.38,178.79) .. (210.84,185.16) .. controls (208.55,170.17) and (198.46,157.75) .. (184.84,152.17) .. controls (185.17,150.05) and (185.33,147.88) .. (185.33,145.67) .. controls (185.33,131.89) and (178.8,119.64) .. (168.67,111.83) .. controls (175.86,106.29) and (184.88,103) .. (194.67,103) -- cycle ;
\draw  [color={rgb, 255:red, 144; green, 19; blue, 254 }  ,draw opacity=1 ][fill={rgb, 255:red, 189; green, 16; blue, 224 }  ,fill opacity=1 ][line width=2.25]  (258,112) .. controls (258,110.34) and (259.34,109) .. (261,109) -- (270.33,109) .. controls (271.99,109) and (273.33,110.34) .. (273.33,112) -- (273.33,121) .. controls (273.33,122.66) and (271.99,124) .. (270.33,124) -- (261,124) .. controls (259.34,124) and (258,122.66) .. (258,121) -- cycle ;
\draw  [color={rgb, 255:red, 74; green, 144; blue, 226 }  ,draw opacity=1 ][fill={rgb, 255:red, 80; green, 227; blue, 194 }  ,fill opacity=1 ][line width=2.25]  (258,133) .. controls (258,131.34) and (259.34,130) .. (261,130) -- (270.33,130) .. controls (271.99,130) and (273.33,131.34) .. (273.33,133) -- (273.33,142) .. controls (273.33,143.66) and (271.99,145) .. (270.33,145) -- (261,145) .. controls (259.34,145) and (258,143.66) .. (258,142) -- cycle ;
\draw  [color={rgb, 255:red, 0; green, 0; blue, 0 }  ,draw opacity=1 ][fill={rgb, 255:red, 0; green, 0; blue, 0 }  ,fill opacity=1 ] (261,202.2) -- (263.33,202.2) -- (263.33,192) -- (268,192) -- (268,202.2) -- (270.33,202.2) -- (265.67,209) -- cycle ;
\draw  [color={rgb, 255:red, 74; green, 144; blue, 226 }  ,draw opacity=1 ][fill={rgb, 255:red, 80; green, 227; blue, 194 }  ,fill opacity=1 ][line width=2.25]  (142.67,103) .. controls (152.45,103) and (161.47,106.29) .. (168.67,111.83) .. controls (158.53,119.64) and (152,131.89) .. (152,145.67) .. controls (152,147.88) and (152.17,150.05) .. (152.49,152.17) .. controls (138.88,157.75) and (128.79,170.17) .. (126.49,185.16) .. controls (110.95,178.79) and (100,163.51) .. (100,145.67) .. controls (100,122.1) and (119.1,103) .. (142.67,103) -- cycle ;
\draw  [color={rgb, 255:red, 144; green, 19; blue, 254 }  ,draw opacity=1 ][fill={rgb, 255:red, 189; green, 16; blue, 224 }  ,fill opacity=1 ][line width=2.25]  (168.67,179.5) .. controls (160.16,172.95) and (154.19,163.26) .. (152.49,152.17) .. controls (157.48,150.13) and (162.94,149) .. (168.67,149) .. controls (174.39,149) and (179.85,150.13) .. (184.84,152.17) .. controls (183.14,163.26) and (177.18,172.95) .. (168.67,179.5) -- cycle ;
\draw  [color={rgb, 255:red, 126; green, 211; blue, 33 }  ,draw opacity=1 ][fill={rgb, 255:red, 184; green, 233; blue, 134 }  ,fill opacity=1 ][line width=2.25]  (258,153) .. controls (258,151.34) and (259.34,150) .. (261,150) -- (270.33,150) .. controls (271.99,150) and (273.33,151.34) .. (273.33,153) -- (273.33,162) .. controls (273.33,163.66) and (271.99,165) .. (270.33,165) -- (261,165) .. controls (259.34,165) and (258,163.66) .. (258,162) -- cycle ;
\draw  [color={rgb, 255:red, 208; green, 2; blue, 27 }  ,draw opacity=1 ][fill={rgb, 255:red, 208; green, 2; blue, 27 }  ,fill opacity=0.63 ][line width=2.25]  (258,173) .. controls (258,171.34) and (259.34,170) .. (261,170) -- (270.33,170) .. controls (271.99,170) and (273.33,171.34) .. (273.33,173) -- (273.33,182) .. controls (273.33,183.66) and (271.99,185) .. (270.33,185) -- (261,185) .. controls (259.34,185) and (258,183.66) .. (258,182) -- cycle ;

\draw (277,171) node [anchor=north west][inner sep=0.75pt]   [align=left] {{\fontfamily{pcr}\selectfont Size }$\displaystyle s_{3}$};
\draw (277,151) node [anchor=north west][inner sep=0.75pt]   [align=left] {{\fontfamily{pcr}\selectfont Size }$\displaystyle s_{2}$};
\draw (98,219) node [anchor=north west][inner sep=0.75pt]   [align=left] {$\displaystyle \oplus K$};
\draw (185,86) node [anchor=north west][inner sep=0.75pt]   [align=left] {$\displaystyle \oplus J$};
\draw (129,86) node [anchor=north west][inner sep=0.75pt]   [align=left] {$\displaystyle \oplus I$};
\draw (247,214) node [anchor=north west][inner sep=0.75pt]   [align=left] {{\fontfamily{pcr}\selectfont Union is }$\displaystyle \oplus I\oplus J\oplus K$};
\draw (277,131) node [anchor=north west][inner sep=0.75pt]   [align=left] {{\fontfamily{pcr}\selectfont Size }$\displaystyle s_{1}$};
\draw (277,111) node [anchor=north west][inner sep=0.75pt]   [align=left] {{\fontfamily{pcr}\selectfont Size }$\displaystyle s_{4}$};

\end{tikzpicture}
\caption{The constraint $\sizeC^s(\oplus I,\oplus J)=1$.}
\label{figure:venn_three_sets}
\end{figure}

We start with the description of $g(s)$. Similar to the previous case, we will fix the sizes of certain sets in the Venn diagram of $\oplus I,\oplus J,\oplus K$ as depicted in~\Cref{figure:venn_three_sets}.
More formally, let $s\in \Nbb^4$. Define $\sizeC^{s}( S_1,S_2,S_3)$ to be the indicator function of 
\[
\abs{S_1\setminus (S_2\cup S_3)}=s_1,\quad \abs{S_2\setminus (S_1\cup S_3)}=s_2,\quad \abs{S_3\setminus (S_1\cup S_2)}=s_3,\quad \abs{ S_1\cap S_2\cap S_3}=s_4.
\] 
Let
\[ 
g(s):=\sum_{\substack{I,J,K\in[n]^t\\ \abs{\oplus I\oplus J\oplus K}=\ell}} \sizeC^{s}( \oplus I, \oplus J,\oplus  K) \cdot \alpha[\oplus I\oplus J\oplus K] \cdot u[I]M_1[I,J]M_2[J,K]v[J].
\] 
We attempt to express $g(s)$ in three different ways as $u^\dagger W_1 R'_1 R'_2 v$, $u^\dagger R_1 W_2 R'_2 v$, and $u^\dagger R_1 R_2 W_3 v$. The simplest to describe is the second expression. Here, we have matrices $R_1,W_2,R_2'$ whose indices are as depicted in~\Cref{figure:decomposition_W_2}.

\begin{figure}[ht]
\centering
\tikzset{every picture/.style={line width=0.75pt}} 

\begin{tikzpicture}[x=0.75pt,y=0.75pt,yscale=-1,xscale=1]

\draw  [line width=1.5]  (7.57,90.67) -- (85.17,90.67) -- (85.17,107.67) -- (7.57,107.67) -- cycle ;
\draw  [fill={rgb, 255:red, 245; green, 166; blue, 35 }  ,fill opacity=1 ][dash pattern={on 0.84pt off 2.51pt}] (106.67,90.67) -- (141.53,90.67) -- (141.53,110.07) -- (106.67,110.07) -- cycle ;
\draw  [color={rgb, 255:red, 0; green, 0; blue, 0 }  ,draw opacity=1 ][fill={rgb, 255:red, 245; green, 166; blue, 35 }  ,fill opacity=1 ][dash pattern={on 0.84pt off 2.51pt}] (141.53,110.07) -- (176.4,110.07) -- (176.4,129.47) -- (141.53,129.47) -- cycle ;
\draw  [fill={rgb, 255:red, 245; green, 166; blue, 35 }  ,fill opacity=1 ][dash pattern={on 0.84pt off 2.51pt}] (176.4,129.47) -- (211.27,129.47) -- (211.27,148.87) -- (176.4,148.87) -- cycle ;
\draw  [fill={rgb, 255:red, 245; green, 166; blue, 35 }  ,fill opacity=1 ][dash pattern={on 0.84pt off 2.51pt}] (211.27,148.87) -- (246.13,148.87) -- (246.13,168.27) -- (211.27,168.27) -- cycle ;
\draw  [fill={rgb, 255:red, 245; green, 166; blue, 35 }  ,fill opacity=1 ][dash pattern={on 0.84pt off 2.51pt}] (483.4,90.67) -- (483.4,125.53) -- (464,125.53) -- (464,90.67) -- cycle ;
\draw  [fill={rgb, 255:red, 245; green, 166; blue, 35 }  ,fill opacity=1 ][dash pattern={on 0.84pt off 2.51pt}] (502.8,125.53) -- (502.8,160.4) -- (483.4,160.4) -- (483.4,125.53) -- cycle ;
\draw  [fill={rgb, 255:red, 245; green, 166; blue, 35 }  ,fill opacity=1 ][dash pattern={on 0.84pt off 2.51pt}] (522.2,160.4) -- (522.2,195.27) -- (502.8,195.27) -- (502.8,160.4) -- cycle ;
\draw  [fill={rgb, 255:red, 245; green, 166; blue, 35 }  ,fill opacity=1 ][dash pattern={on 0.84pt off 2.51pt}] (541.6,195.27) -- (541.6,230.13) -- (522.2,230.13) -- (522.2,195.27) -- cycle ;
\draw  [line width=1.5]  (576.93,90.67) -- (576.93,168.33) -- (559.93,168.33) -- (559.93,90.67) -- cycle ;
\draw  [line width=1.5]  (246.13,90.67) -- (246.13,168.27) -- (106.67,168.27) -- (106.67,90.67) -- cycle ;
\draw  [line width=1.5]  (464,90.67) -- (541.6,90.67) -- (541.6,230.13) -- (464,230.13) -- cycle ;
\draw  [fill={rgb, 255:red, 248; green, 231; blue, 28 }  ,fill opacity=1 ][dash pattern={on 0.84pt off 2.51pt}] (306.53,91.67) -- (306.53,126.53) -- (271.67,126.53) -- (271.67,91.67) -- cycle ;
\draw  [fill={rgb, 255:red, 248; green, 231; blue, 28 }  ,fill opacity=1 ][dash pattern={on 0.84pt off 2.51pt}] (341.4,126.53) -- (341.4,161.4) -- (306.53,161.4) -- (306.53,126.53) -- cycle ;
\draw  [fill={rgb, 255:red, 248; green, 231; blue, 28 }  ,fill opacity=1 ][dash pattern={on 0.84pt off 2.51pt}] (376.37,161.4) -- (376.37,196.27) -- (341.5,196.27) -- (341.5,161.4) -- cycle ;
\draw  [fill={rgb, 255:red, 248; green, 231; blue, 28 }  ,fill opacity=1 ][dash pattern={on 0.84pt off 2.51pt}] (411.23,196.27) -- (411.23,231.13) -- (376.37,231.13) -- (376.37,196.27) -- cycle ;
\draw  [line width=1.5]  (271.67,91.67) -- (411.33,91.67) -- (411.33,231.13) -- (271.67,231.13) -- cycle ;

\draw (42.6,73) node [anchor=north west][inner sep=0.75pt]   [align=left] {$\displaystyle I$};
\draw (92.2,122.2) node [anchor=north west][inner sep=0.75pt]   [align=left] {$\displaystyle I$};
\draw (494,74) node [anchor=north west][inner sep=0.75pt]   [align=left] {$\displaystyle K$};
\draw (414.4,195.95) node [anchor=north west][inner sep=0.75pt]   [align=left] {$\displaystyle J,\oplus K$};
\draw (580.4,120.4) node [anchor=north west][inner sep=0.75pt]   [align=left] {$\displaystyle K$};
\draw (155.4,73) node [anchor=north west][inner sep=0.75pt]   [align=left] {$\displaystyle J,\oplus I$};
\draw (39.8,117) node [anchor=north west][inner sep=0.75pt]   [align=left] {$\displaystyle u$};
\draw (561.6,177.4) node [anchor=north west][inner sep=0.75pt]   [align=left] {$\displaystyle v$};
\draw (165.6,180) node [anchor=north west][inner sep=0.75pt]   [align=left] {$\displaystyle R_{1}$};
\draw (327.6,241.6) node [anchor=north west][inner sep=0.75pt]   [align=left] {$\displaystyle W_{2}$};
\draw (228.4,202.35) node [anchor=north west][inner sep=0.75pt]   [align=left] {$\displaystyle J,\oplus I$};
\draw (321.4,73) node [anchor=north west][inner sep=0.75pt]   [align=left] {$\displaystyle J,\oplus K$};
\draw (490.6,240) node [anchor=north west][inner sep=0.75pt]   [align=left] {$\displaystyle R'_{2}$};

\end{tikzpicture}
\caption{Expressing $g(s)$ as $u^\dagger R_1 W_2R_2' v$.}
\label{figure:decomposition_W_2}
\end{figure}

Based on the intuition from before, there is a very natural way to define these matrices, namely,
\begin{gather*}
R_1[I, (\oplus I', J)]=\indicator\sbra{ \oplus I  = \oplus I'}\cdot M_1[I,J]\quad\text{and}\quad R_2'[(J,\oplus K), K']=\indicator\sbra{ \oplus K' =  \oplus K }\cdot M_2[J,K'],\\
W_2[(J,\oplus I), (J',\oplus K)]=\indicator\sbra{    J=   J'}\cdot \sizeC^{s}(\oplus I,\oplus J,\oplus K) \cdot 
\alpha[\oplus I\oplus J\oplus K].
\end{gather*}
Note here that given the row $(J,\oplus I)$ and the column $(J,\oplus K)$, we can compute $\sizeC^{s}(\oplus I,\oplus J,\oplus K)$ and $\alpha[\oplus I\oplus J\oplus K]$. A similar calculation to before shows that $\|R_1\|,\|R_2'\|\le 1$ and 
\[
\|W_2\| \le \sqrt{{\binom n{s_3}}\cdot{\binom t{s_1}}\binom{t}{s_2}\binom{t}{s_4}}\cdot \sqrt{{\binom n{s_1}}\cdot {\binom t{s_2} \binom{t}{s_3}\binom{t}{s_4}}}= (n/t)^{(s_1+s_3)/2}\cdot t^\ell .
\] 

\begin{figure}[ht]
\centering
\scalebox{0.95}{\input{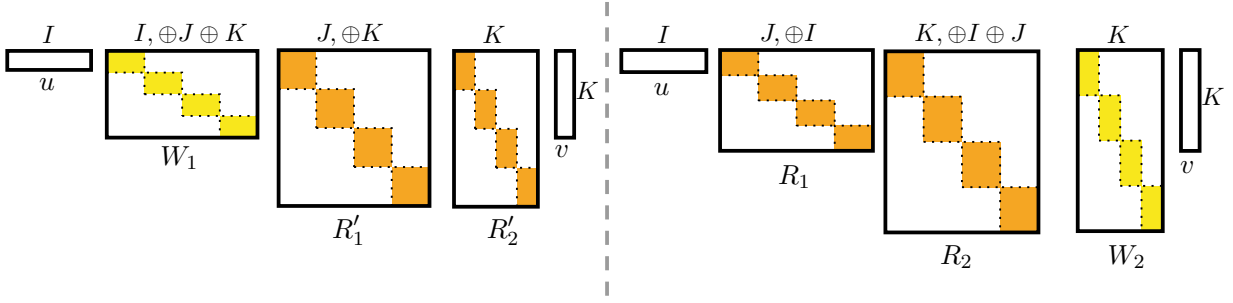}}
\caption{Expressing $g(s)$ as $u^\dagger  W_1 R_1' R_2' v$ and $u^\dagger   R_1 R_2 W_2 v$ respectively.}
\label{figure:decomposition_W_1_2}
\end{figure}

Let us try to define the other two decompositions $u^\dagger W_1R_1'R_2'v$ and $u^\dagger R_1R_2W_2v$ as depicted in~\Cref{figure:decomposition_W_1_2}.
Suppose we could define $W_1$ and  $W_2$ such that
\begin{align}\label{eq:overview_want}
\begin{split} W_1[I,(I',\oplus J\oplus K)]&=\indicator\sbra{ I= I' }\cdot \sizeC^{s}(\oplus I,\oplus J,\oplus K)\cdot 
\alpha[\oplus I\oplus J\oplus K] ,\\
W_3[K,(K',\oplus I\oplus K)]&=\indicator\sbra{  K= K'}\cdot \sizeC^{s}(\oplus I,\oplus J,\oplus K)\cdot
\alpha[\oplus I\oplus J\oplus K].
\end{split}
\end{align}

Then, a calculation similar to the previous case would give the desired operator norm bounds on $W_1$ and $W_3$ as in~\Cref{eq:intro_bound_want}. The problem is that we cannot define matrices $W_1,W_3$ that satisfy~\Cref{eq:overview_want}. 
We explain this issue for $W_1$. Given a row $I$ and a column $(I,\oplus J\oplus K)$, we cannot compute $\sizeC^{s}(\oplus I,\oplus J,\oplus K)$. After all, we only have the information about $\oplus I$ and $\oplus J\oplus K$, and hence the matrix $W_1$ can only enforce the constraints that $\abs{\oplus I\setminus (\oplus J\oplus K)}=s_1+s_4$ and $\abs{(\oplus J\oplus K)\setminus \oplus I}=s_2+s_3$, but it cannot enforce the constraints that $\abs{\oplus J\setminus (\oplus I\cup\oplus K)}=s_2$ or $\abs{\oplus K\setminus (\oplus I\cup\oplus J)}=s_3$. In particular, if we only define $W_1$ to enforce the constraints that it is able to enforce, we will end up counting terms corresponding to $I',J',K'$ which satisfy $\sizeC^{s'}(\oplus I,\oplus J,\oplus K)$ for $s'$ with $s'_2\neq s_2$ and $s'_3\neq s_3$. In this case, instead of estimating the target $g(s)$, we would be over-counting. We need two new ideas here.
\begin{enumerate}
    \item First we need to provide $W_1$ some additional information. One might hope that with a little extra information, $W_1$ can enforce $\sizeC^{s}$, but this turns out to be false. Giving this information will increase the operator norms by too much. Instead, the idea is to provide some information that enforces a variant of the $\sizeC^{s}$ constraint.
    \item This variant will allow us to bound a different function $h(s)$. This function is still an over-counting of $g(s)$, but the important point is that it is a predictable over-counting, that is, $h(s)=\sum_{s'}P[s,s']g(s')$ for some invertible matrix $P$ such that $P^{-1}$ has bounded norm. Therefore, we can derive bounds on $g(s)$ using bounds on $h(s)$.
\end{enumerate}

We first explain step (2). Let $L(I,J,K)=\alpha[\oplus I\oplus J\oplus K]\cdot u[I]M_1[I,J]M_2[J,K]v[K]$. While we would like to bound the expression
\[ 
g(s):=\sum_{I,J,K\in[n]^t}  L(I,J,K)\cdot \sizeC^{s}(I,J,K),
\]
what we can bound turns out to be the expression
\[ 
h(s):=\sum_{I,J,K\in[n]^t\\ } L(I,J,K) \cdot \sum_{\substack{ A,B,C,D\in[n]^t\\A,B,C,D \text{ are disjoint}\\A\cup B\cup C\cup D=\oplus I\oplus J\oplus K}} \subsetC^{s}(A,B,C,D),
\]
where $\subsetC^{s}(A,B,C,D)$ is the indicator function of the constraint that
\begin{equation}\label{eq:subsetC_cond}
 A \subseteq \oplus I , \abs{A}=s_1, \quad
    B\subseteq \oplus J 
    , \abs{B}=s_2,\quad  
    C \subseteq \oplus K
    , \abs{C}=s_3, \quad
    D\subseteq \oplus I\cap \oplus J\cap \oplus K
   , \abs{D}=s_4. 
\end{equation}
This is depicted in~\Cref{figure:comparing_venn_diagrams}. 

\begin{figure}[ht]
\centering
\input{./pics/img8.tex}
\caption{The summation in $g(s)$ versus $h(s)$.}
\label{figure:comparing_venn_diagrams}
\end{figure}

Observe that one of the terms in $h(s)$ is $A=\oplus I\setminus (\oplus J\cup \oplus K),B=\oplus J\setminus (\oplus I\cup \oplus K),C=\oplus K\setminus (\oplus I\cup \oplus J)$ and $D=\oplus I\cap \oplus J\cap\oplus K$. Hence, $h(s)$ consists of $g(s)$ plus some additional terms. 
For example, elements from $D$ can be moved to either $A$, $B$, or $C$ and still satisfy the constraints in \eqref{eq:subsetC_cond}. 
However, we can express 
\[ 
h(s)=\sum_{s'}P[s,s']g(s')
\]
for a structured matrix $P$. This matrix is invertible and has bounded $\|P^{-1}\|_1$. Therefore, our goal of bounding $\|g\|_1$ reduces to bounding $\|h\|_1$ as $h=Pg$. This is done in step (1) which we now explain. 

\begin{figure}[ht]
\centering
\scalebox{0.85}{\input{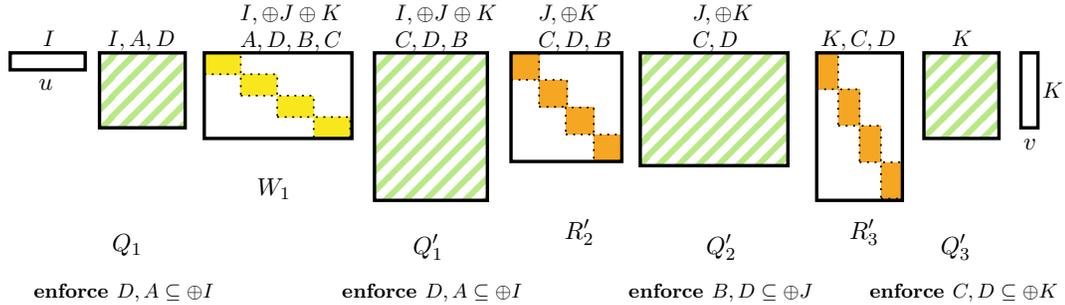}}
\caption{Expressing $h(s)=u^\dagger Q_1W_1Q_1' R_1' Q_2' R_2' Q_3' v$.}
\label{figure:decomposition_Q}
\end{figure}

We now explain how to bound $h(s)$.  We will blow up the matrices in the decomposition $u^\dagger W_1 R_1' R_2' v$ to include information about $A,B,C,D\subseteq[n]$. We will also introduce new matrices $Q_1,Q_1',Q_2',Q_3'$ to enumerate $A,B,C, D$ and verify that they satisfy the $\subsetC^{s}$ constraints in \eqref{eq:subsetC_cond}. Consider the expression $u^\dagger Q_1W_1Q_1' R_1' Q_2' R_2' Q_3' v$, where the matrices are as depicted in~\Cref{figure:decomposition_Q}.

The matrices $W_1,R_2',R_3'$ perform the same role as before and in addition, propagate  information about the sets $A,B,C,D$. The matrices $Q_1, Q_1',Q_2',Q_3'$ impose constraints on $A,B,C,D$ as well as add and delete information as required. In more detail,
\begin{enumerate}
\item $Q_1$ propagates $I$ and introduces $A,D$ such that that $A,D\subseteq \oplus I$, $\abs{A}=s_1$ and $\abs{D}=s_4$.
Given $I$, there are at most ${\binom t{s_1}}\cdot {\binom t{s_4}}$ possibilities for $(A,D)$ and it follows that $\|Q_1\|\le \sqrt{t^{s_1}\cdot t^{s_4}}.$
\item $W_1$ enforces $A\cup B\cup C\cup D=\oplus I\oplus J\oplus K$ and the size constraints on $B,C$.
It also applies $\alpha[\oplus I\oplus J\oplus K]$. For each $I,A,D$, there are at most ${\binom n{s_2}}\cdot {\binom n{s_3}}$ possibilities for $(B,C)$ and once we fix $A,B,C,D$ and $I$, we also fix $\oplus J\oplus K=\oplus I\oplus (A\cup B\cup C\cup D)$. So $\|W_1\|\le \sqrt{n^{s_2}\cdot n^{s_3}}.$
\item $Q_3'$ back-propagates $K$ and introduces $C,D$ such that $C,D\subseteq\oplus K$, $\abs{C}=s_3$, and $\abs{D}=s_4$. Given $K$, there are at most ${\binom t{s_3}}\cdot {\binom t{s_4}}$ possibilities for $(C,D)$ and hence $\|Q_3'\|\le \sqrt{t^{s_3}\cdot t^{s_4}}$.
\item $R_3'$ back-propagates $\oplus K,C,D$ and introduces $J$. It also applies the operator $M_2$. As before, $\|R_3'\|\le 1$. 
\item $Q_2'$ back-propagates $J,\oplus K,C,D$, introduces $B$, and enforces that $B,D\subseteq \oplus J$ and $\abs{B}=s_2$. Given $J$, there are at most ${\binom t{s_2}}$ possibilities for $B$, hence, $\|Q_2'\|\le \sqrt{t^{s_2}}.$
\item $R_2'$ back-propagates  $D,B,C,\oplus J\oplus K$, introduces $I$, and applies the operator $M_1$. As before, $\|R_2'\|\le 1$. 
\item $Q_1'$ back-propagates $D,B,C,\oplus J\oplus K,I$, introduces $A$, and enforces $D,A\subseteq\oplus I$ and $\abs{A}=s_1$. Given $I$, there at most ${\binom t{s_1}}$ possibilities $A$, hence $\|Q_1'\|\le \sqrt{t^{s_1}}$.
\end{enumerate}
Combining all these bounds gives us an upper bound on $h(s)$ of
\[
\sqrt{n^{s_2+s_3}}\cdot t^{s_4+s_1+(s_2+s_3)/2}= (n/t)^{(s_2+s_3)/2}\cdot t^\ell. 
\]
By a symmetric argument, we blow up the matrices
in the decomposition $u^\dagger  R_1 R_2 W_3 v$ to include information about $A,B,C,D\subseteq[n]$ and get
$$h(s) \le  (n/t)^{(s_1+s_2)/2}\cdot t^\ell.$$
Combining the three upper bounds on $h(s)$ we get $h(s) \le (n/t)^{\ell/3} \cdot t^{\ell}$.

\section{Preliminaries}\label{sec:prelim}

For a complex number $x\neq0$, define $\Phase(x)=x/|x|$ as its \emph{phase}; and we additionally define $\Phase(0)=1$.
For sets $S_1,S_2$, we use $S_1\oplus S_2$ to denote their symmetric difference, i.e., $S_1\oplus S_2=\pbra{S_1\setminus S_2}\cup\pbra{S_2\setminus S_1}$.
For a finite set $S$, we use $2^S$ to denote the set of all its subsets.
We use $\Nbb=\cbra{0,1,2,\ldots}$ to denote the set of natural numbers.

\paragraph*{Asymptotics.}
We use $O(\cdot),\Omega(\cdot),\Theta(\cdot)$ to hide universal constants.
$\tilde O(\cdot)$ and $\tilde\Omega(\cdot)$ hide polylogarithmic factors, i.e., $\tilde O(f)=O(f\cdot\polylog f)$ and $\tilde\Omega(f)=\Omega(f\cdot\polylog f)$.
We also use subscript to hide dependence on minor parameters, e.g., $O_{r,d}(f)=O(f\cdot K(r,d))$ for an implicit function $K$.

\paragraph*{Linear Algebra.}
For a (complex) vector, we use $\vabs{\cdot}$ to denote its $\ell_2$-norm; for a (complex) matrix, we use $\vabs{\cdot}$ to denote its operator norm.
We say a vector $u$ is a \emph{unit vector} if $\vabs{u}=1$.
We use $\Ibm_m$ to denote the $m$ by $m$ identity matrix, and, when $m$ is clear from the context, we will simply use $\Ibm$. We use $\Cbb^{[n]\times [m]}$ and $\Cbb^{n\times m}$ to denote the space of complex $n$ by $m$ matrices.

\begin{fact}\label{fct:op_hereditary}
Let $M\in\Cbb^{[n]\times[m]}$ be an $n$ by $m$ matrix. Then for any $S\subseteq[n]$ and $T\subseteq[m]$, we have $\vabs{M[S,T]}\le\vabs{M}$, where $M[S,T]$ is the sub-matrix of $M$ formed by rows in $S$ and columns in $T$.
\end{fact}

\begin{fact}\label{fct:op_block_diagonal}
Assume $M=\diag(M_1,\ldots,M_t)$ is a block diagonal matrix.
Then $\vabs{M}=\max_{i\in [t]}\vabs{M_i}$.
\end{fact}

\begin{fact}[Hölder's Inequality]\label{fct:op_holder}
$\vabs{M}\le\sqrt{\vabs{M}_1\vabs{M}_\infty}$ holds for any matrix $M\in\Cbb^{[n]\times[m]}$, where 
\begin{equation*}
\vabs{M}_1=\max_{1\le j\le m}\sum_{i=1}^n|M[i,j]|
\quad\text{and}\quad
\vabs{M}_\infty=\max_{1\le i\le n}\sum_{j=1}^m|M[i,j]|.
\end{equation*}
\end{fact}

\paragraph*{Parallel Quantum Queries.}
We use the following standard model for quantum query algorithms with parallel queries \cite{DBLP:journals/ipl/Montanaro10}.
Let $O_x$ be the standard quantum query oracle of input $x=x_1\cdots x_n\in\binpm^n$.
That is, $O_x$ acts on an $(n+1)$-dimensional space indexed by basis states $\ket{0},\ket{1},\ldots,\ket{n}$, and performs the operation $O_x\ket{0}=\ket{0}$ and $O_x\ket{i}=x_i\ket{i}$ for each $i\in[n]$.

Let $\Acal$ be a quantum query algorithm that makes $r$ rounds of adaptive queries with $t$ parallel queries per round.
Assume it uses $w$ auxiliary qubits, and fails to compute a Boolean function $f\colon\binpm^n\to\binpm$ with probability at most $\eps$, then it is equivalent to the existence of 
\begin{itemize}
\item a unit state $\ket{\psi}\in\Cbb^{\cbra{0,\ldots,n}^t\times[2^w]}$,
\item $r-1$ unitary matrices $U_1,\ldots,U_{r-1}$,
\item a measurement matrix $M$ that $\vabs{M}\le1$ and $M$ is positive semi-definite,
\end{itemize}
such that
$$
\vabs{\sqrt M(O_x^{\otimes t}\otimes\Ibm_{2^w})U_{r-1}\cdots U_2(O_x^{\otimes t}\otimes\Ibm_{2^w})U_1(O_x^{\otimes t}\otimes\Ibm_{2^w})\ket\psi}^2
\begin{cases}
\ge1-\eps&\text{ for all $x\in f^{-1}(1)$,}\\
\le\eps&\text{ for all $x\in f^{-1}(0)$.}
\end{cases}
$$

We remark that another natural way of describing the quantum query is through an oracle $O_x'$, which acts on a $2(n+1)$-dimensional space indexed by basis states $\cbra{\ket{i}\ket{b}}_{i\in\cbra{0,\ldots,n},b\in\binpm}$, and performs the operation $O_x'\ket{0}\ket{b}=\ket{0}\ket{b}$ and $O_x'\ket{i}\ket{b}=\ket{i}\ket{b\cdot x_i}$ for each $i\in[n],b\in\binpm$. 
It turns out that the two models are equivalent \cite{DBLP:journals/eatcs/HoyerS05,DBLP:journals/ipl/Montanaro10} in the sense that
$$
O_x'=V_1\pbra{O_x\otimes\Ibm_2}V_2
$$
for some unitary matrices $V_1,V_2$.
We will use the standard model with the $O_x$ oracle, which is more convenient for our purposes.
\section{Fourier Growth of the Quantum Query Model}\label{sec:fourier_growth_same}

One way to think of $O_x$ is to view the input $x$ as a truth table of length $(n+1)$ where $x_0$ is fixed to $1$. In this sense, the oracle query can be unified as $O_x\ket{i}=x_i\ket{i}$ for all $i\in\cbra{0,\ldots,n}$.
Meanwhile, for our purposes, it is desired to obtain Fourier growth bounds for downwards closed families. That is, the Fourier growth bounds should hold for the function even after fixing variables to values.
This is usually not an issue regarding complexity measures, but the quantum query model is not evidently downwards closed. 
Therefore we prove the following more general theorem.

\begin{theorem}\label{thm:fourier_growth_same}
Let $n,t\ge1$ and $m\ge0$ and $d\ge2$ be integers. Define $A=[n]^t\times[m]$.
Let $u,v\in\Cbb^A$ be two unit vectors. 
Let $M_1,\ldots,M_{d-1}\in\Cbb^{A\times A}$ be matrices satisfying $\vabs{M_i}\le1$ for each $i\in[d-1]$.
Define $f\colon\binpm^{[n]}\to\Cbb$ as
$$
f(x)
=u^\dag\pbra{O_x^{\otimes t}\otimes\Ibm_m}M_1\pbra{O_x^{\otimes t}\otimes \Ibm_m}M_2\cdots M_{d-1}\pbra{O_x^{\otimes t}\otimes\Ibm_m}v.
$$

Let $\rho\in\cbra{\pm1,*}^{[n]}$ be an arbitrary restriction\footnote{$f|_\rho$ is a sub-function on $\tilde n$ variables of $f$ by fixing $x_i$ to $\rho(i)$ for $i\notin\rho^{-1}(*)$.} and $\tilde n=\abs{\rho^{-1}(*)}$.
Then for any $\ell\ge0$, we have
$$
L_{1,\ell}(f|_\rho)
=\sum_{S\subseteq\rho^{-1}(*),|S|=\ell}\abs{\hat{f|_\rho}(S)}
\le2^{\kappa(d,\ell)}\cdot t^\ell\cdot\max\cbra{1,(\tilde n/t)^{\frac12\floorbra{(d-1)\ell/d}}},
$$
where $\kappa(d,\ell)=O(d\ell)\cdot\min\cbra{2^{d\ell},\ell^{2^d}}$.
\end{theorem}

Before proving \Cref{thm:fourier_growth_same}, we first summarize its application to the Fourier growth of quantum query algorithms, via the conversion stated in \Cref{sec:prelim}.

\begin{corollary}[Formal Version of \Cref{thm:fourier_growth}]\label{cor:foureri_growth_quantum_query}
Assume $\Acal$ is a query algorithm given oracle access $O_x$ and uses arbitrarily many auxiliary qubits.
Assume $\Acal$ makes $r$ rounds of queries where each round consists of $t\le n$ parallel queries.
Let $f\colon\binpm^{[n]}\to[0,1]$ be its acceptance probability, i.e., $f(x)=\Pr\sbra{\Acal\text{ accepts x}}$.
Then
$$
L_{1,\ell}(f)\le2^{\kappa(r,\ell)}\cdot t^\ell\cdot(n/t)^{\frac12\floorbra{\frac{(2r-1)\ell}{2r}}},
$$
where $\kappa(r,\ell)=O(r\ell)\cdot\min\cbra{2^{2r\ell},\ell^{4^r}}$.

Moreover, this bound holds when some bits of $x$ are fixed in advance.
\end{corollary}

Now we proceed to the proof of \Cref{thm:fourier_growth_same}.
Note that
$$
f(x)
=\sum_{\substack{\alpha_1,\ldots,\alpha_d\in[m]\\I_1,\ldots,I_d\in[n]^t}}
\pbra{
u[(I_1,\alpha_1)]
v[(I_d,\alpha_d)]
\prod_{i\in[d-1]}M_i[(I_i,\alpha_i),(I_{i+1},\alpha_{i+1})]}
\cdot
\prod_{j\in[d],k\in[t]}x[I_j(k)].
$$
By rearranging coordinates, we assume without loss of generality $\rho^{-1}(*)=[\tilde n]$, i.e., $\rho$ fixes all but the first $\tilde n$ bits of $x$.
Now we expand $f|_\rho$ as
\begin{align*}
f|_\rho(x)
=\sum_{\substack{\alpha_1,\ldots,\alpha_d\in[m]\\I_1,\ldots,I_d\in[n]^t}}
&\pbra{
u[(I_1,\alpha_1)]
v[(I_d,\alpha_d)]
\prod_{i\in[d-1]}M_i[(I_i,\alpha_i),(I_{i+1},\alpha_{i+1})]}\\
&\quad\cdot
\pbra{\prod_{\substack{j\in[d],k\in[t]\\I_j(k)>\tilde n}}\rho[I_j(k)]}
\pbra{\prod_{\substack{j\in[d],k\in[t]\\I_j(k)\le\tilde n}}x[I_j(k)]}.
\end{align*}
Recall that $A=[n]^t\times[m]$ is the space of parallel queries and ancillary qubits.
Now we define matrices $\tilde M_1,\ldots,\tilde M_{d-1}\in\Cbb^{A\times A}$ as
$$
\tilde M_i[(I_i,\alpha_i),(I_{i+1},\alpha_{i+1})]=M_i[(I_i,\alpha_i),(I_{i+1},\alpha_{i+1})]\cdot\prod_{k\in[t],I_i(k)>\tilde n}\rho[I_i(k)]
$$
then define vectors $\tilde u,\tilde v\in\Cbb^A$ as $\tilde u=u$ and
$$
\tilde v[(I_d,\alpha_d)]=v[(I_d,\alpha_d)]\cdot\prod_{k\in[t],I_d(k)>\tilde n}\rho[I_d(k)].
$$
Therefore we have
\begin{align*}
f|_\rho(x)
=\sum_{\substack{\alpha_1,\ldots,\alpha_d\in[m]\\I_1,\ldots,I_d\in[n]^t}}
&\pbra{
\tilde u[(I_1,\alpha_1)]
\tilde v[(I_d,\alpha_d)]
\prod_{i\in[d-1]}\tilde M_i[(I_i,\alpha_i),(I_{i+1},\alpha_{i+1})]}
\cdot
\prod_{\substack{j\in[d],k\in[t]\\I_j(k)\le\tilde n}}x[I_j(k)].
\end{align*}
In addition, each $\tilde M_i,\tilde u,\tilde v$ is the original $M_i,u,v$ left multiplied by a $\pm1$-diagonal matrix.
By the norm guarantees of $M_i,u,v$, this means
\begin{equation}\label{eq:thm:fourier_growth_same_1}
\vabs{\tilde u}=\vabs{\tilde v}=1
\quad\text{and}\quad
\vabs{\tilde M_i}\le1.
\end{equation}

Note that any index appearing twice in the multi-set $\cbra{I_j(k)}_{j\in[d],k\in[t]}$ cancels due to $(\pm1)^2=1$.
We can compute each Fourier coefficient $\hat{f|_\rho}(S)$ as
\begin{align*}
\hat{f|_\rho}(S)
&=\sum_{\substack{(I_1,\alpha_1),\ldots,(I_d,\alpha_d)\in A\\\oplus I_1\oplus\cdots\oplus I_d=S}}
\tilde u[(I_1,\alpha_1)]
\tilde v[(I_d,\alpha_d)]
\prod_{i\in[d-1]}\tilde M_i[(I_i,\alpha_i),(I_{i+1},\alpha_{i+1})],
\end{align*}
where, from now on, we use $\oplus T_1\oplus T_2\oplus\cdots\subseteq[\tilde n]$ to denote the set of indices in $[\tilde n]$ that appear odd times in the multi-set consisting of indices from $T_1,T_2,\ldots$.

Now we introduce $a(S)=\Phase^{-1}\pbra{\hat{f|_\rho}(S)}$ to denote the inverse of the phase of $\hat{f|_\rho}(S)$, then
\begin{align}
L_{1,\ell}(f|_\rho)
&=\sum_{S\subseteq[\tilde n],|S|=\ell}a(S)\cdot\hat{f|_\rho}(S)\notag\\
&=\sum_{\substack{(I_1,\alpha_1),\ldots,(I_d,\alpha_d)\in A\\\abs{\oplus I_1\oplus\cdots\oplus I_d}=\ell}}
\underbrace{a(\oplus I_1\oplus\cdots\oplus I_d)\cdot
\tilde u[(I_1,\alpha_1)]
\tilde v[(I_d,\alpha_d)]
\prod_{i\in[d-1]}\tilde M_i[(I_i,\alpha_i),(I_{i+1},\alpha_{i+1})]
}_{L(I_1,\alpha_1,\ldots,I_d,\alpha_d)}.
\label{eq:thm:fourier_growth_same_2}
\end{align}

For the analysis purpose, we will partition the binary strings $\bin^d$ and for this we introduce some notation.
For each $b,b'\in\bin^d$, we say $b'\ge b$ if $b'$ is no smaller than $b$ entrywise; and we say $b'>b$ if $b'\ge b$ and $b'\neq b$.
For $i\in[d]$, we write $i\in b$ if $b_i=1$, and $i\notin b$ if $b_i=0$.

Let $B=\cbra{b\in\bin^d\mid\vabs{b}_1\equiv1\mod2}$ be the set of strings of odd Hamming weights, according to which we will partition $\oplus I_1\oplus\cdots\oplus I_d$ into parts based on the membership in each $\oplus I_i$. 
Formally, for each $I=(I_1,\ldots,I_d)\in\pbra{[n]^t}^d$ and $b\in B$, define $I^{(b)}\subseteq[\tilde n]$ to be the set of indices in $[\tilde n]$ that appears in and only in those $\oplus I_i$ satisfying $b_i=1$.
Formally,
$$
I^{(b)}=\pbra{\bigcap_{i\in b}\oplus I_i}\setminus\pbra{\bigcup_{i\notin b}\oplus I_i}.
$$
We emphasize that the intersection $\cap$ and union $\cup$ operators are applied to the inner sets $\oplus I_i$.
Due to the construction, $|B|=2^{d-1}$ and $\oplus I_1\oplus\cdots\oplus I_d$ equals the (disjoint) union of all $I^{(b)}$'s.

Recall the definition of $L(I_1,\alpha_1,\ldots,I_d,\alpha_d)$ from \Cref{eq:thm:fourier_growth_same_2}.
Now for each $s=(s^{(b)})_{b\in B}\in\Nbb^B$ satisfying $\vabs{s}_1=\ell$, we write the contribution of all the $(I_1,\alpha_1,\ldots,I_d,\alpha_d)$ consistent with $s$ as
\begin{equation}\label{eq:thm:fourier_growth_same_3}
g(s)=\sum_{\substack{(I_1,\alpha_1),\ldots,(I_d,\alpha_d)\in A\\\abs{I^{(b)}}=s^{(b)},\forall b\in B}}
L(I_1,\alpha_1,\ldots,I_d,\alpha_d).
\end{equation}
Then we can express $L_{1,\ell}(f|_\rho)$ equivalently as 
\begin{equation}\label{eq:thm:fourier_growth_same_4}
L_{1,\ell}(f|_\rho)=\sum_{s\in\Nbb^B,\vabs{s}_1=\ell}g(s).
\end{equation}
Here we are grouping $(I_1,\ldots,I_d)$ based on the sizes of the intersections and bounding the contribution from each group separately. We now count the number of possible sizes of intersection patterns. By a balls-into-bins counting, there are only
\begin{equation}\label{eq:thm:fourier_growth_same_6}
D:=\binom{\ell+|B|-1}{|B|-1}=\binom{\ell+2^{d-1}-1}{2^{d-1}-1}=O_{d,\ell}(1)
\end{equation}
many possible $s$ in the summation of \Cref{eq:thm:fourier_growth_same_4}. 

Thus, our goal becomes bounding $\|g\|_1=\sum_{\|s\|_1=\ell}\abs{g(s)}$ and to do this we would like to bound each $g(s)$. However, as described in the proof overview, what we can bound turns out to be a function $h(s)$ where $h(s)=\sum_{\|s'\|_1=\ell} g(s')\cdot P[s,s']$ for some matrix $P$. We will now describe this function $h$ and the matrix $P$. \Cref{lem:fourier_growth_same_h_bound} will prove an upper bound on each $\abs{h(s)}$ and we will use this lemma and properties about $P$ to show the desired bound on $\|g\|_1$.

For each $s=(s^{(b)})_{b\in B}\in\Nbb^B$ satisfying $\vabs{s}_1=\ell$, define
\begin{equation}\label{eq:thm:fourier_growth_same_5}
h(s)=
\sum_{\substack{(I_1,\alpha_1),\ldots,(I_d,\alpha_d)\in A\\\abs{\oplus I_1\oplus\cdots\oplus I_d}=\ell}}
\sum_{\substack{J^{(b)}\subseteq[\tilde n]\text{ of size }s^{(b)},\forall b\in B\\\text{$J^{(b)}$'s are pairwise disjoint}\\J^{(b)}\subseteq\bigcup_{b'\ge b}I^{(b')},\forall b\in B}}
L(I_1,\alpha_1,\ldots,I_d,\alpha_d).
\end{equation}
See \Cref{sec:overview} for a concrete example for the relation between $h$ and $g$.

Each $h(\cdot)$ will be reformulated as a product of matrices that we can bound.
\begin{lemma}\label{lem:fourier_growth_same_h_bound}
$|h(s)|\le t^\ell\cdot\max\cbra{1,(\tilde n/t)^{\frac12\floorbra{(d-1)\ell/d}}}$.
\end{lemma}

The proof of \Cref{lem:fourier_growth_same_h_bound} is deferred to the end of this section. Now we continue the task of bounding $L_{1,\ell}(f|_\rho)$ assuming \Cref{lem:fourier_growth_same_h_bound}.
To relate $h(\cdot)$ with $g(\cdot)$, we count for any fixed $(I_1,\alpha_1),\ldots,(I_d,\alpha_d)\in A$ satisfying $\abs{\oplus I_1\oplus\cdots\oplus I_d}=\ell$, the number of possible $(J^{(b)})_{b\in B}$.
By the condition $J^{(b)}\subseteq\bigcup_{b'\ge b}I^{(b')},\forall b\in B$, we enumerate $J^{(b)}$ in the decreasing order of $\vabs{b}_1,b\in B$.
Then the number of possibilities for each $J^{(b)}$ is exactly 
$$
\binom{\sum_{b'\ge b}\abs{I^{(b')}}-\sum_{b'>b}\abs{J^{(b')}}}{\abs{J^{(b)}}}=
\binom{\sum_{b'\ge b}\abs{I^{(b')}}-\sum_{b'>b}s^{(b')}}{s^{(b)}},
$$
where we fix $s=(s^{(b)})_{b\in B}$ and each $J^{(b)}$ has size $s^{(b)}$.
Therefore the total number of choices is the telescoping product
$$
\prod_{b\in B}\binom{\sum_{b'\ge b}\abs{I^{(b')}}-\sum_{b'>b}s^{(b')}}{s^{(b)}},
$$
which allows us to rewrite $h(s)$ as
\begin{align}
h(s)
&=\sum_{\substack{(I_1,\alpha_1),\ldots,(I_d,\alpha_d)\in A\\\abs{\oplus I_1\oplus\cdots\oplus I_d}=\ell}} L(I_1,\alpha_1,\ldots,I_d,\alpha_d)\cdot\prod_{b\in B}\binom{\sum_{b'\ge b}\abs{I^{(b')}}-\sum_{b'>b}s^{(b')}}{s^{(b)}}
\tag{recall \Cref{eq:thm:fourier_growth_same_5}}\\
&=\sum_{s'\in\Nbb^B,\vabs{s'}_1=\ell}
\prod_{b\in B}\binom{\sum_{b'\ge b}{s'}^{(b')}-\sum_{b'>b}s^{(b')}}{s^{(b)}}\cdot
\sum_{\substack{(I_1,\alpha_1),\ldots,(I_d,\alpha_d)\in A\\\abs{I^{(b')}}={s'}^{(b')},\forall b'\in B}}
L(I_1,\alpha_1,\ldots,I_d,\alpha_d)
\notag\\
&=\sum_{s'\in\Nbb^B,\vabs{s'}_1=\ell}
g(s')\cdot
\prod_{b\in B}\binom{\sum_{b'\ge b}{s'}^{(b')}-\sum_{b'>b}s^{(b')}}{s^{(b)}}
\tag{recall \Cref{eq:thm:fourier_growth_same_3}}\\
&=:\sum_{s'\in\Nbb^B,\vabs{s'}_1=\ell}
g(s')\cdot P[s,s'].
\label{eq:thm:fourier_growth_same_7}
\end{align}
Therefore, viewing $h$ and $g$ as two vectors, they satisfy the relation $h=Pg$ where $P$ is the coefficient matrix defined above and the dimension of $P$ is $D \times D$ by \Cref{eq:thm:fourier_growth_same_6}.
Recall the $\ell_1$ norm of a matrix from \Cref{fct:op_holder}.
The following lemma studies the properties of $P$ itself. 

\begin{lemma}\label{lem:fourier_growth_same_P_bound}
$P$ is an invertible matrix over $\Cbb$ and $\vabs{P^{-1}}_1\le D\cdot\binom{\ell\cdot2^{d-1}}{\ell}^D$.
\end{lemma}

Then with the same linear algebraic notation, \Cref{eq:thm:fourier_growth_same_4} completes the proof:
\begin{align*}
L_{1,\ell}(f|_\rho)
&=(1^D)^\top g\le\vabs{g}_1=\vabs{P^{-1}h}_1\le\vabs{P^{-1}}_1\vabs{h}_1
\le D\cdot\binom{\ell\cdot2^{d-1}}{\ell}^D\cdot\vabs{h}_1
\tag{by \Cref{lem:fourier_growth_same_P_bound}}\\
&\le D\cdot\binom{\ell\cdot2^{d-1}}{\ell}^D\cdot D\cdot\vabs{h}_\infty\\
&\le D^2\cdot\binom{\ell\cdot2^{d-1}}{\ell}^D\cdot t^\ell\cdot\max\cbra{1,(\tilde n/t)^{\frac12\floorbra{(d-1)\ell/d}}}
\tag{by \Cref{lem:fourier_growth_same_h_bound}}\\
&\le2^{O(d\ell D)}\cdot t^\ell\cdot\max\cbra{1,(\tilde n/t)^{\frac12\floorbra{(d-1)\ell/d}}}.
\end{align*}
Finally by \Cref{eq:thm:fourier_growth_same_6}, we note that
$$
D=\begin{cases}
\binom{\ell+2^{d-1}-1}\ell\le\binom{2^d}\ell\le2^{d\ell} & \ell\le2^{d-1}-1,\\
\binom{\ell+2^{d-1}-1}{2^{d-1}-1}\le\binom{2\ell}{2^{d-1}-1}\le(2\ell)^{2^{d-1}-1}\le\ell^{2^d} & \ell\ge2^{d-1}.
\end{cases}
$$
This gives the desired bounds in \Cref{cor:foureri_growth_quantum_query}.

Now we prove \Cref{lem:fourier_growth_same_P_bound}.
\begin{proof}[Proof of \Cref{lem:fourier_growth_same_P_bound}]
We extend the partial order $>$ on elements in $B$ to an arbitrary total order, denoted as $\gg$.\footnote{For example, one can think of $\gg$ as the decreasing order in the Hamming weight, and lexicographical order within the same Hamming weight.}
Let $T=\cbra{s\in\Nbb^B\mid\vabs{s}_1=\ell}$ and let $\ggg$ be the lexicographical order on $T$ induced by $\gg$ on $B$, i.e., $s\ggg s'$ iff there exists some $b\in B$ such that $s^{(b)}>{s'}^{(b)}$ and $s^{(b')}={s'}^{(b')}$ holds for all $b'\gg b$.

Now we show that the matrix $P$, with rows and columns sorted according to $\ggg$, is lower-triangular with ones on the diagonal.
This proves that $P$ is invertible and $\det(P)=1$.
Let $s\ggg s'$ be arbitrary elements from $T$ and let $b^*\in B$ be such that $s^{(b^*)}>{s'}^{(b^*)}$ and $s^{(b')}={s'}^{(b')}$ holds for all $b'\gg b^*$.
Then we have
\begin{align*}
P[s,s']
&=\prod_{b\in B}\binom{\sum_{b'\ge b}{s'}^{(b')}-\sum_{b'>b}s^{(b')}}{s^{(b)}}
\tag{recall \Cref{eq:thm:fourier_growth_same_7}}\\
&=\binom{{s'}^{(b^*)}+\sum_{b'>b^*}{s'}^{(b')}-\sum_{b'>b^*}s^{(b')}}{s^{(b^*)}}
\cdot\prod_{b\neq b^*}\binom{\sum_{b'\ge b}{s'}^{(b')}-\sum_{b'>b}s^{(b')}}{s^{(b)}}\\
&=\binom{{s'}^{(b^*)}+\sum_{b'\gg b^*}{s'}^{(b')}-\sum_{b'\gg b^*}s^{(b')}}{s^{(b^*)}}
\cdot\prod_{b\neq b^*}\binom{\sum_{b'\ge b}{s'}^{(b')}-\sum_{b'>b}s^{(b')}}{s^{(b)}}
\tag{since $\gg$ is extended from $>$}\\
&=0
\cdot\prod_{b\neq b^*}\binom{\sum_{b'\ge b}{s'}^{(b')}-\sum_{b'>b}s^{(b')}}{s^{(b)}}
=0.
\tag{since $s^{(b^*)}>{s'}^{(b^*)}$ and $s^{(b')}={s'}^{(b')}$ for all $b'\gg b^*$}
\end{align*}
The diagonal values can be calculated similarly:
$$
P[s,s]
=\prod_{b\in B}\binom{\sum_{b'\ge b}s^{(b')}-\sum_{b'>b}s^{(b')}}{s^{(b)}}
=\prod_{b\in B}\binom{s^{(b)}}{s^{(b)}}
=1.
$$

Now we bound $\vabs{P^{-1}}_1$.
Let $P_{-s,-s'}$ be matrix $P$ removing the $s$-th row and the $s'$-th column.
Then the matrix inversion formula (See e.g., \cite{wiki:Adjugate_matrix}) gives 
$$
\abs{P^{-1}[s,s']}
=\abs{\frac{\det\pbra{P_{-s',-s}}}{\det(P)}}
=\abs{\det\pbra{P_{-s',-s}}}
\le\per(P)
\le\vabs{P}_1^D,
$$
where $\per(\cdot)$ denotes the permanent and the last inequality uses the fact that the dimension of $P$ is $D$.
Thus 
\begin{equation}\label{eq:fourier_growth_same_P_bound_1}
\vabs{P^{-1}}_1\le D\cdot\max_{s,s'\in T}\abs{P^{-1}[s,s']}\le D\cdot \vabs{P}_1^D
\end{equation}
and it suffices to bound $\vabs{P}_1=\max_{s'\in T}\sum_{s\in T}|P[s,s']|$.
Fix the maximizer $s'$, we have
\begin{align*}
\vabs{P}_1
&=\sum_{s\in T}\prod_{b\in B}\binom{\sum_{b'\ge b}{s'}^{(b')}-\sum_{b'>b}s^{(b')}}{s^{(b)}}
\le\sum_{s\in T}\prod_{b\in B}\binom{\sum_{b'\ge b}{s'}^{(b')}}{s^{(b)}}\\
&\le\sum_{s\in T}\prod_{b\in B}\binom{\ell}{s^{(b)}}
\tag{since $\vabs{s'}_1=\ell$}\\
&=\text{the coefficient of $x^\ell$ in }\pbra{(1+x)^\ell}^{|B|}
\tag{since $\vabs{s}_1=\ell$}\\
&=\binom{\ell\cdot2^{d-1}}{\ell},
\tag{since $|B|=2^{d-1}$}
\end{align*}
which completes the proof by plugging into \Cref{eq:fourier_growth_same_P_bound_1}.
\end{proof}

Finally we prove \Cref{lem:fourier_growth_same_h_bound} which bounds $h(\cdot)$ entrywise.
For convenience, we recall the definition of $h(s)$ and $L(I_1,\alpha_1,\ldots,I_d,\alpha_d)$ from \Cref{eq:thm:fourier_growth_same_5} and \Cref{eq:thm:fourier_growth_same_2}:
$$
h(s)=
\sum_{\substack{(I_1,\alpha_1),\ldots,(I_d,\alpha_d)\in A\\\abs{\oplus I_1\oplus\cdots\oplus I_d}=\ell}}
\sum_{\substack{J^{(b)}\subseteq[\tilde n]\text{ of size }s^{(b)},\forall b\in B\\\text{$J^{(b)}$'s are pairwise disjoint}\\J^{(b)}\subseteq\bigcup_{b'\ge b}I^{(b')},\forall b\in B}}
L(I_1,\alpha_1,\ldots,I_d,\alpha_d)
$$
and
$$
L(I_1,\alpha_1,\ldots,I_d,\alpha_d)=a(\oplus I_1\oplus\cdots\oplus I_d)\cdot
\tilde u[(I_1,\alpha_1)]
\tilde v[(I_d,\alpha_d)]
\prod_{i\in[d-1]}\tilde M_i[(I_i,\alpha_i),(I_{i+1},\alpha_{i+1})].
$$

\begin{proof}[Proof of \Cref{lem:fourier_growth_same_h_bound}]
Let $r\in[d]$ be an index to be optimized later.
We will write $h(s)$ as a product of matrices:
\begin{equation}\label{eq:lem:fourier_growth_same_h_bound}
h(s)=\bar u^\dag Q_1 R_1 Q_2 R_2 \cdots Q_{r-1} R_{r-1} Q_r W Q_r' R'_r Q_{r+1}' R'_{r+1} \cdots Q_{d-1}' R'_{d-1} Q_d' \bar v,
\end{equation}
where 
\begin{itemize}
\item $Q_i$ enforces constraints on $J^{(b)}$s and propagates information about them forward,
\item $Q_i'$ enforces constraints on $J^{(b)}$s and propagates information about them backward,
\item $R_i$  implements the action of $\tilde M_i$ and propogates information about $\oplus I_1\oplus\cdots\oplus I_i$ forward,
\item $R_i'$ implements the action of $\tilde M_i$ and propogates information about $\oplus I_{i+1}\oplus\cdots\oplus I_{d-1}$ backward,
\item $W$ is a sign matrix constructed to multiply by the phases $a(\cdot)$, as well as aggregate information about $J^{(b)},\oplus I_1\oplus \ldots \oplus I_i$ and $\oplus I_{i+1}\oplus \ldots \oplus I_{d-1}$ .
\item the vector $\bar u$ (resp., $\bar v$) is simply the vector $u$ (resp., $v$) padded with zeros to fit with the dimension of $Q_1$ (resp., $Q_d'$).
\end{itemize}
In the following, we use the symbol $\bot$ to denote the value is unassigned. 
For any $b\in\bin^d$ and $i\in[d]$, we use $b_{\le i}$ to denote string $(b_1,b_2,\ldots,b_i)$, and define similarly for $b_{<i},b_{\ge i},b_{>i}$.

We index the rows of matrix $\matsym\in\cbra{Q_i,R_i,W,Q_i',R_i'}$ by $(I_\matsym,\alpha_\matsym)\in A$, $S_\matsym\subseteq[\tilde n]$, and $J_\matsym^{(b)}\in2^{[\tilde n]}\cup\cbra{\bot}$ for all $b\in B$; and its columns are similarly indexed by $(I_\matsym',\alpha_\matsym')$, $S_\matsym'$, and ${J_\matsym'}^{(b)}$.

Likewise, we index the coordinates of vector $\vecsym\in\cbra{\bar u,\bar v}$ by $(I_\vecsym,\alpha_\vecsym)$, $S_\vecsym$, and $J_\vecsym^{(b)}$.
In particular for the vectors, we assign
$$
\vecsym[(I_\vecsym,\alpha_\vecsym,S_\vecsym,J_\vecsym^{(b)})]
=\begin{cases}
\vecsym[(I_\vecsym,\alpha_\vecsym)] & S_\vecsym=\emptyset\text{ and } J_\vecsym^{(b)}=\bot,\forall b\in B,\\
0 & \text{otherwise}.
\end{cases}
$$
Despite the dimension of the vectors $\bar u,\bar v$ being increased, they are simply padded by zeros. Therefore the norm is preserved from \Cref{eq:thm:fourier_growth_same_1}:
\begin{equation}\label{eq:baruv_norm}
\vabs{\bar u}=\vabs{\bar v}=1.
\end{equation}

Now we turn to the matrices.

\paragraph*{Construction of the Blow-Up Matrix $Q_i$.}
Each $\matsym=Q_i$ is a zero-one matrix where the entry is assigned one iff $I_\matsym'=I_\matsym$, $\alpha_\matsym'=\alpha_\matsym$, $S_\matsym'=S_\matsym$, and for each $b\in B$,

\begin{enumerate}
\item\label{itm:Qi_1} 
if $i\in b$, then ${J_\matsym'}^{(b)}\subseteq\oplus I_\matsym$,
\item 
\begin{enumerate}
    \item\label{itm:Qi_2} If $b_{\le i}=0^{i-1}1$, then $J_\matsym^{(b)}=\bot$ and $\abs{{J_\matsym'}^{(b)}}=s^{(b)}$,
     \item\label{itm:Qi_3}
    If $b_{\le i}\neq 0^{i-1}1$, then $J_\matsym^{(b)}={J_\matsym'}^{(b)}$,
\end{enumerate}
\end{enumerate}

The intuition behind this expression is the following. We need to ensure two conditions, namely, (1) for all $i$, we have $J^{(b)}\subseteq \oplus I_i$ if $ b\ni i$, and, (2) $\abs{J^{(b)}}=s^{(b)}$ for all $b$. Condition (1) will be checked by the matrix $Q_i$ in~\Cref{itm:Qi_1}. Condition (2) will be checked by the matrix $Q_i$ in~\Cref{itm:Qi_2}, where $i$ is the first non-zero coordinate in $b$.
In contrast, \Cref{itm:Qi_3} is to inherent Condition (2) from previous blow-up matrices.
It will turn out that these conditions are enough to guarantee that the sets $J^{(b)}$ are pairwise disjoint (as shown in~\Cref{eq:h_bound_almost_1} and~\Cref{eq:h_bound_almost_2}).

We now upper bound the operator norm of $Q_i$.
On the one hand, each column of $\matsym$ has at most one non-zero entry since the row index is a refinement of the column index.
On the other hand, each row of $\matsym$ only has the possible freedom to select ${J_\matsym'}^{(b)}$ if $b_{\le i}=0^{i-1}1$ in \Cref{itm:Qi_2}, each of which amounts to at most $\binom{\abs{\oplus I_\matsym}}{s^{(b)}}\le\binom t{s^{(b)}}$ options.
Therefore by \Cref{fct:op_holder}, we have
\begin{equation}\label{eq:Qi_norm}
\vabs{Q_i}
\le\sqrt{\prod_{b:b_{\le i}=0^{i-1}1}\binom t{s^{(b)}}}.
\end{equation}

\paragraph*{Construction of the Blow-Up Matrix $Q'_i$.}
Each $\matsym=Q_i'$ is a zero-one blow-up matrix similar to $Q_i$, with the role of the columns and rows exchanged: The entry is assigned one iff $I_\matsym=I_\matsym'$, $\alpha_\matsym=\alpha_\matsym'$, $S_\matsym=S_\matsym'$, and for each $b\in B$,
\begin{enumerate}
\item
if $i\in b$, then ${J_\matsym}^{(b)}\subseteq\oplus I_\matsym$, and
\item \begin{enumerate}
    \item \label{itm:Qi'_2}  if $b_{\ge i}=10^{d-i}$, then ${J_\matsym'}^{(b)}=\bot$ and $\abs{{J_\matsym}^{(b)}}=s^{(b)}$,
    \item\label{itm:Qi'_1} if $b_{\ge i}\neq 10^{d-i}$, then ${J_\matsym'}^{(b)}=J_\matsym^{(b)}$.
\end{enumerate}
\end{enumerate}
By the same argument for \Cref{eq:Qi_norm}, we have
\begin{equation}\label{eq:Qi'_norm}
\vabs{Q_i'}
\le\sqrt{\prod_{b:b_{\ge i}=10^{d-i}}\binom t{s^{(b)}}}.
\end{equation}

\paragraph*{Construction of the Operator Matrices $R_i,R_i'$.}
Each $\matsym=R_i$ is constructed to implement the action of $\tilde M_i$ as well as to propagate forward information about $\oplus I_1\oplus\cdots\oplus I_i$ and this information will be captured by $S_\matsym'$.
Each entry of $R_i$ is either $\tilde M_i[(I_\matsym,\alpha_\matsym),(I'_\matsym,\alpha_\matsym')]$ or zero, where the former case requires $S_\matsym'=\oplus S_\matsym\oplus I_\matsym$ and ${J_\matsym'}^{(b)}=J_\matsym^{(b)}$ for all $b\in B$.

To bound its operator norm, we view the row index $(I_\matsym,\alpha_\matsym,S_\matsym,\{J_\matsym^{(b)}\})$ as $(I_\matsym,\alpha_\matsym,T_\matsym,\{J_\matsym^{(b)}\})$ where $T_\matsym=\oplus S_\matsym\oplus I_\matsym$. 
Note that this is indeed a bijection since $S_\matsym=\oplus T_\matsym\oplus I_\matsym$.
Moreover in the new indexing way, the entry is $\tilde M_i[(I_\matsym,\alpha_\matsym),(I'_\matsym,\alpha_\matsym)]$ iff $S_\matsym'=T_\matsym$ and ${J_\matsym'}^{(b)}= J_\matsym^{(b)}$, which means $\matsym=R_i$ is a block diagonal matrix with block indexed by $(T_\matsym,\{J_\matsym^{(b)}\})$.
Since each block is a sub-matrix of $\tilde M_i$, by \Cref{fct:op_block_diagonal} and \Cref{fct:op_hereditary} the operator norm is preserved from \Cref{eq:thm:fourier_growth_same_1}:
\begin{equation}\label{eq:Ri_norm}
\vabs{R_i}\le1.
\end{equation}

Each $\matsym=R_i'$ is similarly constructed to implement $\widetilde{M}_i$, as well as to propagate information about $\oplus I_{i+1}\oplus\cdots\oplus I_{d-1}$ using $S_\matsym$: Its entry is either $\tilde M_i[(I_\matsym,\alpha_\matsym),(I_\matsym',\alpha_\matsym')]$ or zero, where the former case requires $S_\matsym=\oplus S_\matsym'\oplus I_\matsym'$ and $J_\matsym^{(b)}={J_\matsym'}^{(b)}$ for all $b\in B$.
By the same argument, we have
\begin{equation}\label{eq:Ri'_norm}
\vabs{R_i'}\le1.
\end{equation}

\paragraph*{Construction of the Sign Matrix $W$.}
The final piece is to incorporate phases $a(\cdot)$ in the matrix $\matsym=W$.
To this end, the entry is assigned $a(\oplus S_\matsym\oplus I_\matsym\oplus S_\matsym')$ if (otherwise the entry is assigned zero)
\begin{enumerate}
\item\label{itm:W} 
$I_\matsym=I_\matsym'$, $\alpha_\matsym=\alpha_\matsym'$, and $\abs{\oplus S_\matsym\oplus I_\matsym\oplus S'_\matsym}=\ell$,
\item for each $b\in B$,
\begin{enumerate}
\item\label{itm:W_1} 
if $b_{\le r}=0^r$, then $J_\matsym^{(b)}=\bot$, ${J'_\matsym}^{(b)}\neq\bot$, and $\abs{{J'_\matsym}^{(b)}}=s^{(b)}$,
\item\label{itm:W_2} 
else if $b_{\ge r}=0^{d-r+1}$, then ${J_\matsym'}^{(b)}=\bot$, $J_\matsym^{(b)}\neq\bot$, and $\abs{J_\matsym^{(b)}}=s^{(b)}$,
\item\label{itm:W_3} 
else (i.e., $b_{\le r}\neq0^r$ and $b_{\ge r}\neq0^{d-r+1}$), then $J_\matsym^{(b)}={J_\matsym'}^{(b)}\subseteq[\tilde n]$ of size $s^{(b)}$,
\end{enumerate}
\item\label{itm:W_4} 
$\oplus S_\matsym\oplus I_\matsym\oplus S'_\matsym=\bigcup_{b:b_{\le r}=0^r}{J'}^{(b)}\cup\bigcup_{b:b_{\le r}\neq0^r}J^{(b)}$.
\end{enumerate}
The analysis of $\vabs{W}$ is similar to the one of $\vabs{Q_i}$.
Each row of $\matsym$ is allowed to select ${J_\matsym'}^{(b)}$ if $b_{\le r}=0^r$ in \Cref{itm:W_1}, each of which has at most $\binom{\tilde n}{s^{(b)}}$ options.
Let $\bar S=\bigcup_{b:b_{\le r}=0^r}{J'}^{(b)}\cup\bigcup_{b:b_{\le r}\neq0^r}J^{(b)}$, which is fixed after enumerating ${J_\matsym'}^{(b)}$'s.
By \Cref{itm:W_4}, we have $S'_\matsym=\oplus S_\matsym\oplus I_\matsym\oplus\bar S$ which is also fixed.
Since each $a(\cdot)$ is a phase which has unit norm, we have
$$
\vabs{W}_\infty
\le\prod_{b:b_{\le r}=0^r}\binom{\tilde n}{s^{(b)}}.
$$
Similarly, we can bound $\vabs{W}_1\le\prod_{b:b_{\ge r}=0^{d-r+1}}\binom{\tilde n}{s^{(b)}}$.
Therefore by \Cref{fct:op_holder}, we have
\begin{equation}\label{eq:W_norm}
\vabs{W}\le\sqrt{
\prod_{b:b_{\le r}=0^r}\binom{\tilde n}{s^{(b)}}\cdot\prod_{b:b_{\ge r}=0^{d-r+1}}\binom{\tilde n}{s^{(b)}}}.
\end{equation}

\paragraph*{Optimizing Bounds.}
To conclude the proof of \Cref{lem:fourier_growth_same_h_bound}, it suffices to verify \Cref{eq:lem:fourier_growth_same_h_bound} and optimize the choice of $r\in[d]$.
We will deal with the former later, and focus on the bounds first.

Assuming \Cref{eq:lem:fourier_growth_same_h_bound}, we have
\begin{align*}
\abs{h(s)}
&\le
\vabs{\bar u}\vabs{Q_1}\vabs{R_1}\cdots\vabs{Q_{r-1}}\vabs{R_{r-1}}\vabs{Q_r}\vabs{W}\vabs{Q_r'}\vabs{R'_r}\cdots\vabs{Q_{d-1}'}\vabs{R'_{d-1}}\vabs{Q_d'}\vabs{\bar v}\\
&\le
\sqrt{
\prod_{i=1}^r\prod_{b:b_{\le i}=0^{i-1}1}\binom t{s^{(b)}}
\cdot
\prod_{i=r}^d\prod_{b:b_{\ge i}=10^{d-i}}\binom t{s^{(b)}}
\cdot
\prod_{b:b_{\le r}=0^r}\binom{\tilde n}{s^{(b)}}\cdot\prod_{b:b_{\ge r}=0^{d-r+1}}\binom{\tilde n}{s^{(b)}}
}
\tag{by \Cref{eq:baruv_norm}, \Cref{eq:Qi_norm}, \Cref{eq:Qi'_norm}, \Cref{eq:Ri_norm}, \Cref{eq:Ri'_norm}, and \Cref{eq:W_norm}}\\
&=
\sqrt{
\prod_{b:b_{\le r}\neq0^r}\binom t{s^{(b)}}
\cdot
\prod_{b:b_{\ge r}\neq0^{d-r+1}}\binom t{s^{(b)}}
\cdot
\prod_{b:b_{\le r}=0^r}\binom{\tilde n}{s^{(b)}}
\cdot
\prod_{b:b_{\ge r}=0^{d-r+1}}\binom{\tilde n}{s^{(b)}}
}\\
&\le
(\sqrt t)^{\sum_{b:b_{\le r}\neq0^r}s^{(b)}+\sum_{b:b_{\ge r}\neq0^{d-r+1}}s^{(b)}}
(\sqrt{\tilde n})^{\sum_{b:b_{\le r}=0^r}s^{(b)}+\sum_{b:b_{\ge r}=0^{d-r+1}}s^{(b)}}\\
&=t^\ell\cdot
(\sqrt{\tilde n/t})^{\sum_{b:b_{\le r}=0^r}s^{(b)}+\sum_{b:b_{\ge r}=0^{d-r+1}}s^{(b)}}
\tag{since $\vabs{s}_1=\ell$}\\
&=:t^\ell\cdot
(\sqrt{\tilde n/t})^{e_r}.
\end{align*}

If $t\ge\tilde n$, then $|h(s)|\le t^\ell$ since $e_r\ge0$.
Now consider the case $t\le\tilde n$.
For each $b\in B$, define 
$$
z(b)=\max\cbra{r\in\Nbb\mid j\notin b,\forall j\le r}
\quad\text{and}\quad
z'(b)=\min\cbra{r\in\Nbb\mid j\notin b,\forall j\ge r}.
$$
Since $\vabs{b}_1\ge1$, we have $0\le z(b)<d$, $0<z'(b)\le d+1$, and $z(b)\le z'(b)-2$.
Therefore
\begin{align*}
\sum_{r=1}^de_r
&=\sum_{r=1}^d\sum_{b:b_{\le r}=0^r}s^{(b)}+\sum_{r=1}^d\sum_{b:b_{\ge r}=0^{d-r+1}}s^{(b)}
=\sum_{b\in B}s^{(b)}\cdot\pbra{z(b)+d+1-z'(b)}\\
&\le\sum_{b\in B}s^{(b)}\cdot(d-1)=(d-1)\ell,
\tag{since $z(b)\le z'(b)-2$ and $\vabs{s}_1=\ell$}
\end{align*}
which by averaging argument implies there exists a choice $r\in[d]$ such that $e_r\le\floorbra{(d-1)\ell/d}$.
This particular choice of $r$ allows us to bound
$$
|h(s)|\le t^\ell\cdot(\tilde n/t)^{\frac12\floorbra{(d-1)\ell/d}}
$$
as desired.

\paragraph*{Verifying \Cref{eq:lem:fourier_growth_same_h_bound}.}
Finally we verify the multiplication in \Cref{eq:lem:fourier_growth_same_h_bound} is consistent with the definition of $h(s)$ in \Cref{eq:thm:fourier_growth_same_5}.

For each $i\in[r]$, define vector $y^{(i)}=\pbra{\bar u^\dag Q_1R_1Q_2\cdots R_{i-1}Q_i}^\dag$. For any fixed $I_1,\ldots,I_i$, define the following indicator functions.
\[\sizelC^{(i)}(\{J^{(b)}\}):=\indicator\sbra{\begin{array}{ll} 
{J}^{(b)}=\bot & \text{if }b_{\le i}=0^{i}\\
\abs{{J}^{(b)}}=s^{(b)} &\text{if } b_{\le i}\neq0^{i}
\end{array}}\]
\[\subsetlC^{(i)}(\{J^{(b)}\}):=\indicator\sbra{ \oplus I_j \supseteq J^{(b)},\forall j\le i,b\ni j}\]
We also define an indicator functions that captures the constraints of $Q_i$.
\[\QC^{(i)}(\{J^{(b)},{J'}^{(b)}\}):= \indicator\sbra{\begin{array}{ll} 
{J}^{(b)}=\bot,{J'}^{(b)}\subseteq\oplus I\text{ has size }s^{(b)} 
& b_{\le i}=0^{i-1}1\\
{J'}^{(b)}={J}^{(b)}\subseteq\oplus I & b_{\le i-1}\neq0^{i-1},b_i=1\\
{J'}^{(b)}={J}^{(b)} & b_i=0
\end{array}}. \]
\begin{claim} The $(I,\alpha,S,\{J^{(b)}\})$-th entry of $y^{(i)}$ equals 
\begin{equation}\label{eq:yi_induction_alt}
\sum_{\substack{(I_1,\alpha_1),\ldots,(I_i,\alpha_i)\in A\\I_i=I,\alpha_i=\alpha\\ \oplus I_1\oplus\cdots\oplus I_{i-1}=S}} \sizelC^{(i)}(\{J^{(b)}\})\cdot \subsetlC^{(i)}(\{J^{(b)}\})\cdot 
\tilde u[(I_1,\alpha_1)]\prod_{j=1}^{i-1}\tilde M_j[(I_j,\alpha_j),(I_{j+1},\alpha_{j+1})] 
\end{equation} \label{claim:expression}
\end{claim}
We proof~\Cref{claim:expression} by induction. We now verify the base case $i=1$. In this case, the expression in~\Cref{claim:expression} reduces to $\indicator\sbra{\begin{array}{ll}  J^{(b)}=\bot &\text{if } b_1=0\\ \abs{J^{(b)}}=s^{(b)} &\text{if }b_1=1\end{array}}\cdot \underset{\oplus I_1\supseteq J^{(b)},\forall b\ni 1}{\sum} \widetilde{u}[(I,\alpha)]$. We have $y^{(1)}=\pbra{\bar u^\dag Q_1}^\dag$, and the $(I,\alpha,S,\{J^{(b)}\})$-th entry of $y^{(1)}$ equals
\begin{align*}
&=\sum_{(I',\alpha',S',\{{J'}^{(b)}\})}\bar u[(I',\alpha',S',\{{J'}^{(b)}\})]\cdot Q_1[(I',\alpha',S',\{{J'}^{(b)}\}),(I,\alpha,S,\{{J}^{(b)}\})]\\
&=\sum_{(I',\alpha',\emptyset,\{\bot\})}u[(I',\alpha')]\cdot Q_1[(I',\alpha',\emptyset,\{\bot\}),(I,\alpha,S,\{{J}^{(b)}\})]
\tag{by the definition of $\bar u$}\\
&=u[(I,\alpha)]\cdot\indicator[S=\bot]\cdot \indicator
\sbra{
\begin{array}{ll} 
J^{(b)}\subseteq\oplus I, \abs{J^{(b)}}=s^{(b)} & \text{if }b\ni1,\\
=\bot & \text{if }b\not\ni1
\end{array}}.
\tag{by the definition of $Q_1$}
\end{align*} 
This proves the base case of $i=1$ for~\Cref{claim:expression}. We now handle the inductive case $i\ge2$. We have $y^{(i)}=\pbra{\pbra{y^{(i-1)}}^\dag R_{i-1}Q_i}^\dag$ and hence, the $(I,\alpha,S,\{J^{(b)}\})$-th entry of $y^{(i)}$ equals the sum over all possible $(I',\alpha',S',\{{J'}^{(b)}\})$ and $(I'',\alpha'',S'',\{{J''}^{(b)}\})$ of the product of the following three terms:
\begin{enumerate}
    \item $y^{(i-1)}[(I',\alpha',S',\{{J'}^{(b)}\})]$,
    \item $R_{i-1}[(I',\alpha',S',\{{J'}^{(b)}\}),(I'',\alpha'',S'',\{{J''}^{(b)}\})]$, and 
    \item $Q_i[(I'',\alpha'',S'',\{{J''}^{(b)}\}),(I,\alpha,S,\{J^{(b)}\})]$. 
\end{enumerate}
It is easy to see from the definition of $R_{i-1}$ that (2) is non-zero only if $S''=\oplus I'\oplus S'$ and ${J''}^{(b)}={J'}^{(b)}$.
Similarly, it follows from the definition of $Q_i$ that (3) is non-zero only if $I''=I$, $\alpha''=\alpha$, and $S''=S$. We now use the inductive hypothesis to express the $(I,\alpha,S,\{J^{(b)}\})$-th entry of $y^{(i)}$ as the sum over all possible $(I',\alpha',S',\{{J'}^{(b)}\})$ where $S=\oplus I'\oplus S'$ of the product of the following three terms:
\begin{enumerate}
    \item $y^{(i-1)}[(I',\alpha',S',\{{J'}^{(b)}\})]$, which by induction can be expressed as
\begin{align*}
\sum_{\substack{(I_1,\alpha_1),\ldots,(I_{i-1},\alpha_{i-1})\in A\\I_{i-1}=I',\alpha_{i-1}=\alpha'\\ \oplus I_1\oplus\cdots\oplus I_{i-2}=S'}} 
&\sizelC^{(i-1)}(\{{J'}^{(b)}\})\cdot \subsetlC^{(i-1)}(\{{J'}^{(b)}\})\\
&\qquad\cdot
\tilde u[(I_1,\alpha_1)]\prod_{j=1}^{i-2}\tilde M_j[(I_j,\alpha_j),(I_{j+1},\alpha_{j+1})].
\end{align*}
    \item $R_{i-1}[(I',\alpha',S',\{{J'}^{(b)}\}),(I,\alpha,S,\{{J'}^{(b)}\})]$, which is equal to $\tilde M_{i-1}[(I',\alpha'),(I,\alpha)]$.
    \item $Q_i[(I,\alpha,S,\{{J'}^{(b)}\}),(I,\alpha,S,\{J^{(b)}\})]$, which is equal to $\QC^{(i-1)}(\{{J'}^{(b)},J^{(b)}\})$
\end{enumerate}
We now combine the indicator functions in (1) and (3) by a case analysis. It is not too difficult to show that
\begin{align*}
&\sizelC^{(i-1)}(\{{J'}^{(b)}\})\cdot \QC^{(i-1)}(\{{J'}^{(b)},J^{(b)}\})\\
&=\sizelC^{(i)}(\{{J}^{(b)}\})\cdot \indicator\sbra{ \begin{array}{ll} {J'}^{(b)}={J}^{(b)}=\bot & \text{if }b_{\le i}=0^i\\ 
{J'}^{(b)}=\bot,J^{(b)}\subseteq \oplus I& \text{if }b_{\le i}=0^{i-1}1\\ 
{J'}^{(b)}={J}^{(b)}\subseteq \oplus I& \text{if }b_{\le i}\neq 0^{i-1},b_i=1\\ 
{J'}^{(b)}={J}^{(b)} & \text{if }b_{\le i}\neq 0^{i-1},b_i=0\\ 
\end{array}}.
\end{align*}
Furthermore, setting $I_i=I$, we have
\begin{align*}
&\indicator\sbra{ \begin{array}{ll} {J'}^{(b)}={J}^{(b)}=\bot & \text{if }b_{\le i}=0^i\\ 
{J'}^{(b)}=\bot,J^{(b)}\subseteq \oplus I& \text{if }b_{\le i}=0^{i-1}1\\ 
{J'}^{(b)}={J}^{(b)}\subseteq \oplus I& \text{if }b_{\le i}\neq 0^{i-1},b_i=1\\ 
{J'}^{(b)}={J}^{(b)} & \text{if }b_{\le i}\neq 0^{i-1},b_i=0\\ 
\end{array}}\cdot \subsetlC^{(i-1)}(\{{J'}^{(b)}\})=\subsetlC^{(i)}(\{J^{(b)}\}).
\end{align*}
Putting this together with the above facts completes the proof of~\Cref{claim:expression}.

The $(I,\alpha,S,\{J^{(b)}\})$-th entry of ${y'}^{(i)}:=Q_i'R_i'\cdots Q_{d-1}'R_{d-1}'Q_d'\bar v$ can be analyzed analogously as
\begin{equation}\label{eq:yi'_induction}
\sum_{\substack{(I_i,\alpha_i),\ldots,(I_d,\alpha_d)\in A\\I_i=I,\alpha_i=\alpha\\ \oplus I_{i+1}\oplus\cdots\oplus I_d=S}} \sizegC^{(i)}(\{J^{(b)}\}) \cdot \subsetgC^{(i)}(\{J^{(b)}\})\cdot 
\tilde v[(I_d,\alpha_d)]\prod_{j=i}^{d-1}\tilde M_j[(I_j,\alpha_j),(I_{j+1},\alpha_{j+1})].
\end{equation}
where any fixed $I_1,\ldots,I_i$, we define the following indicator functions.
\[\sizegC^{(i)}(\{J^{(b)}\}):=\indicator\sbra{\begin{array}{ll} 
{J}^{(b)}=\bot & \text{if }b_{\ge i}=0^{d-i+1}\\
\abs{{J}^{(b)}}=s^{(b)} &\text{if } b_{\ge i}\neq0^{d-i+1}
\end{array}}\]
\[\subsetgC^{(i)}(\{J^{(b)}\}):=\indicator\sbra{ \oplus I_j \supseteq J^{(b)},\forall j\ge i,b\ni j}\]
Hence the RHS of \Cref{eq:lem:fourier_growth_same_h_bound} equals $\pbra{y^{(r)}}^\dag W{y'}^{(r)}$ and evaluates to 
\begin{align}
&\sum_{\substack{(I,\alpha,S,\{J^{(b)}\})\\(I',\alpha',S',\{{J'}^{(b)}\})}}
y^{(r)}[(I,\alpha,S,\{J^{(b)}\})]
W[(I,\alpha,S,\{J^{(b)}\}),(I',\alpha',S',\{{J'}^{(b)}\})]
{y'}^{(r)}[(I',\alpha',S',\{{J'}^{(b)}\})]
\notag\\
&=\sum_{\substack{I,\alpha,S,S'\\\{J^{(b)}\},\{{J'}^{(b)}\}\\\abs{\oplus S\oplus I\oplus S'}=\ell}}
y^{(r)}[(I,\alpha,S,\{J^{(b)}\})]
W[(I,\alpha,S,\{J^{(b)}\}),(I,\alpha,S',\{{J'}^{(b)}\})]
{y'}^{(r)}[(I,\alpha,S',\{{J'}^{(b)}\})]
\tag{by condition \Cref{itm:W} of the definition of $W$}\\
&=\sum_{\substack{I,\alpha,S,S'\\\{J^{(b)}\},\{{J'}^{(b)}\}\\\abs{\oplus S\oplus I\oplus S'}=\ell}}
W[(I,\alpha,S,\{J^{(b)}\}),(I,\alpha,S',\{{J'}^{(b)}\})]
\notag\\
&\hspace{20pt}\cdot
\sum_{\substack{(I_1,\alpha_1),\ldots,(I_r,\alpha_r)\in A\\I_r=I,\alpha_r=\alpha\\ \oplus I_1\oplus\cdots\oplus I_{r-1}=S}}\sizelC^{(r)}(\{J^{(b)}\})\cdot \subsetlC^{(r)}(\{J^{(b)}\})
\cdot
\tilde u[(I_1,\alpha_1)]\prod_{i=1}^{r-1}\tilde M_i[(I_i,\alpha_i),(I_{i+1},\alpha_{i+1})]
\tag{by \Cref{claim:expression}}\\
&\hspace{20pt}\cdot
\sum_{\substack{(I_r,\alpha_r),\ldots,(I_d,\alpha_d)\in A\\I_r=I,\alpha_r=\alpha\\ \oplus I_{r+1}\oplus\cdots\oplus I_d=S'}} \sizegC^{(r)}(\{{J'}^{(b)}\})\cdot \subsetgC^{(r)}(\{{J'}^{(b)}\})
\cdot
\tilde v[(I_d,\alpha_d)]\prod_{i=r}^{d-1}\tilde M_i[(I_i,\alpha_i),(I_{i+1},\alpha_{i+1})]
\tag{by \Cref{eq:yi'_induction}}\\
&=\sum_{\substack{(I_1,\alpha_1),\ldots,(I_d,\alpha_d)\in A\\\abs{\oplus I_1\oplus\cdots\oplus I_d}=\ell}}
\tilde u[(I_1,\alpha_1)]\tilde v[(I_d,\alpha_d)]\prod_{i=1}^{d-1}\tilde M_i[(I_i,\alpha_i),(I_{i+1},\alpha_{i+1})]
\label{eq:yWy'}\\
&\hspace{20pt}
\cdot\sum_{\{J^{(b)}\},\{{J'}^{(b)}\}}
W[(I_r,\alpha_r,\oplus I_1\oplus\cdots\oplus I_{r-1},\{J^{(b)}\}),(I_r,\alpha_r,\oplus I_{r+1}\oplus\cdots\oplus I_d,\{{J'}^{(b)}\})]
\label{eq:yWy'_0}\\
&\hspace{20pt}\cdot \sizelC^{(r)}(\{J^{(b)}\})\cdot \subsetlC^{(r)}(\{J^{(b)}\})\cdot \sizegC^{(r)}(\{{J'}^{(b)}\})\cdot \subsetgC^{(r)}(\{{J'}^{(b)}\}).
\label{eq:yWy'_1}
\end{align}
Notice that $b_{\le r}$ and $b_{\ge r}$ cannot both be zeros for $b\in B$.
Thus conditions \Cref{itm:W_1,itm:W_2,itm:W_3} of $W$ show that we can enumerate ${J''}^{(b)}\subseteq[\tilde n]$ of size $s^{(b)}$ and then let $J^{(b)},{J'}^{(b)}$ be $\bot$ or ${J''}^{(b)}$ based on $b$.
After this, \Cref{eq:yWy'_1} simply becomes the indicator of ${J''}^{(b)}\subseteq\bigcap_{i\in b}\oplus I_i$, and condition \Cref{itm:W_4} of $W$ becomes $\oplus I_1\oplus\cdots\oplus I_d=\bigcup_{b\in B}{J''}^{(b)}$.
That is, \Cref{eq:yWy'_0} and \Cref{eq:yWy'_1} when combined, is equal to
$$
a(\oplus I_1\oplus\cdots\oplus I_d)\cdot\sum_{\substack{{J''}^{(b)}\subseteq[\tilde n]\text{ of size }s^{(b)},\forall b\in B\\{J''}^{(b)}\subseteq\bigcap_{i\in b}\oplus I_i}}
\sbra{\oplus I_1\oplus\cdots\oplus I_d=\bigcup_{b\in B}{J''}^{(b)}}.
$$
Now recall the definition of $L(I_1,\alpha_1,\ldots,L_d,\alpha_d)$ from \Cref{eq:thm:fourier_growth_same_2} and combine \Cref{eq:yWy'}.
The RHS of \Cref{eq:lem:fourier_growth_same_h_bound} equals
\begin{equation}\label{eq:h_bound_almost_1}
\sum_{\substack{(I_1,\alpha_1),\ldots,(I_d,\alpha_d)\in A\\\abs{\oplus I_1\oplus\cdots\oplus I_d}=\ell\\{J''}^{(b)}\subseteq[\tilde n]\text{ of size }s^{(b)}}}
\sbra{{J''}^{(b)}\subseteq\bigcap_{i\in b}\oplus I_i
\quad\land\quad
\oplus I_1\oplus\cdots\oplus I_d=\bigcup_{b\in B}{J''}^{(b)}}
\cdot L(I_1,\alpha_1,\ldots,I_d,\alpha_d).
\end{equation}
Finally it suffices to show this is equivalent to the summation in \Cref{eq:thm:fourier_growth_same_5} which we restate here:
\begin{equation}\label{eq:h_bound_almost_2}
\sum_{\substack{(I_1,\alpha_1),\ldots,(I_d,\alpha_d)\in A\\\abs{\oplus I_1\oplus\cdots\oplus I_d}=\ell\\J^{(b)}\subseteq[\tilde n]\text{ of size }s^{(b)}}}
\sbra{J^{(b)}\subseteq\bigcup_{b'\ge b}I^{(b')}
\quad\land\quad
\text{$J^{(b)}$'s are pairwise disjoint}
}
\cdot L(I_1,\alpha_1,\ldots,I_d,\alpha_d),
\end{equation}
where we recall that $I^{(b')}=\pbra{\bigcap_{i\in b'}\oplus I_i}\setminus\pbra{\bigcup_{i\notin b'}\oplus I_i}$.
To this end, we fix $(I_1,\alpha_1),\ldots,(I_d,\alpha_d)\in A$ satisfying $\abs{\oplus I_1\oplus\cdots\oplus I_d}=\ell$ and show each possible $\{{J''}^{(b)}\}$ from \Cref{eq:h_bound_almost_1} is also counted as $\{J^{(b)}\}$ in \Cref{eq:h_bound_almost_2}, and vice versa.

\indent\textbf{From \Cref{eq:h_bound_almost_1} to \Cref{eq:h_bound_almost_2}.}
By the definition of $I^{(b')}$, we know $\oplus I_i\cap I^{(b')}=\emptyset$ whenever $i\notin b'$.
Since ${J''}^{(b)}\subseteq\bigcap_{i\in b}\oplus I_i$, we have ${J''}^{(b)}\cap I^{(b')}\neq\emptyset$ implies $b'\ge b$.
Note that $\oplus I_1\oplus\cdots\oplus I_d=\bigcup_bI^{(b)}$.
Therefore $\bigcup_b{J''}^{(b)}=\bigcup_bJ^{(b)}$, and thus ${J''}^{(b)}\subseteq\bigcup_{b'\ge b}I^{(b')}$ as desired in \Cref{eq:h_bound_almost_2}.
On the other hand, $\sum_b|{J''}^{(b)}|=\vabs{s}_1=\ell=\abs{\oplus I_1\oplus\cdots\oplus I_d}$. 
Thus $\oplus I_1\oplus\cdots\oplus I_d=\bigcup_b{J''}^{(b)}$ implies that ${J''}^{(b)}$'s are pairwise disjoint as desired in \Cref{eq:h_bound_almost_2}.

\indent\textbf{From \Cref{eq:h_bound_almost_2} to \Cref{eq:h_bound_almost_1}.}
Since $\oplus I_1\oplus\cdots\oplus I_d=\bigcup_bI^{(b)}$, we have $\bigcup_bJ^{(b)}\subseteq\bigcup_b\bigcup_{b'\ge b}I^{(b')}=\bigcup_bI^{(b')}=\oplus I_1\oplus\cdots\oplus I_d$.
On the other hand, $\sum_b|J^{(b)}|=\vabs{s}_1=\ell=\abs{\oplus I_1\oplus\cdots\oplus I_d}$. 
Thus $J^{(b)}$'s being pairwise disjoint implies that $\oplus I_1\oplus\cdots\oplus I_d=\bigcup_bJ^{(b)}$ as desired in \Cref{eq:h_bound_almost_1}.
By the definition of $I^{(b')}$, we know $I^{(b')}\subseteq\bigcap_{i\in b'}\oplus I_i$.
Therefore $J^{(b)}\subseteq\bigcup_{b'\ge b}I^{(b')}\subseteq\bigcup_{b'\ge b}\bigcap_{i\in b'}\oplus I_i=\bigcap_{i\in b}\oplus I_i$ as desired in \Cref{eq:h_bound_almost_1}.
\end{proof}

\section*{Acknowledgement}
We thank anonymous QIP'24 reviewers for helpful comments.
KW also wants to thank Guangxu Yang and Penghui Yao for references in the quantum communication complexity.

\bibliographystyle{alpha} 
\bibliography{ref}

\appendix
\section[Proof of Theorem 1.1]{Proof of \Cref{thm:separation}}\label{app:thm:separation}

Before proving \Cref{thm:separation}, we first define the $k$-fold Forrelation problem.

\begin{definition}[$k$-fold Forrelation Problem]\label{def:kforr}
For an integer $k\ge 2$, the $k$-fold Forrelation problem is a partial Boolean function on $n$ bits. 
Let $H = H_n$ denote the $n\times n$ (orthonormal) Hadamard matrix where $n=2^m$ for $m \in \Nbb$. 
Let $x_1, \ldots, x_k \in \binpm^{n}$ denote truth tables of $k$ different Boolean functions. 
Define the degree-$k$ polynomial $\kForr : \binpm^{kn} \to \Rbb$ as follows 
\begin{equation}\label{eqn:forr}
    \ \kForr(x) = \frac{1}{n} \sum_{(i_1,\ldots,i_k) \in [n]^k} x_1({i_1}) \cdot H_{i_1,i_2} \cdot x_2({i_2}) \cdot H_{i_2,i_3}\cdots \cdot x_{k-1}(i_{k-1})\cdot H_{i_{k-1},i_{k}}\cdot x_k({i_k}).
\end{equation}
The $k$-fold Forrelation problem is to decide whether $|\kForr(x)|\le \frac\delta2$ or  $\kForr(x)\ge \delta$ for a parameter $\delta$. For the applications in this paper, we take $\delta=2^{-5k}$.
\end{definition}

\begin{fact}[{\cite{aaronson2015forrelation}}]\label{fct:kforr_alg}
There exists a quantum circuit $Q$ that makes $\ceilbra{k/2}$ queries and uses $O(k\log n)$ gates, such that for any input $x\in\binpm^{kn}$, it holds that
$$
\Pr\sbra{Q\text{ accepts }x}=\frac{1+\kForr(x)}2.
$$
\end{fact}

Now we proceed to the proof of \Cref{thm:separation}.
Since \Cref{itm:thm:separation_1} is proved in \cite{aaronson2015forrelation} using \Cref{fct:kforr_alg}, here we only do the calculation for \Cref{itm:thm:separation_2}.
By \cite[Theorem 3.2 and Theorem 3.4]{bansal2021k}, it suffices to show that
\begin{equation}\label{eq:reduction_to_bs}
\pbra{\frac1{\sqrt n}}^{1-\frac1{2r}}\cdot\pbra{L_{1,\ell}(f)}^{1/\ell}\le r^{-20}
\quad\forall~2r\le\ell\le 2r\cdot(2r-1)
\text{ and }
t\le O_r\pbra{n^{c(r,r')}}
\end{equation}
where the Fourier growth bound $L_{1,\ell}(f)$ is from \Cref{thm:fourier_growth} satisfies
$$
L_{1,\ell}(f)
\le O_{r',\ell}\pbra{t^\ell\cdot \left(\sqrt{n/t}\right)^{\floorbra{ \left(1-\frac{1}{2r'}\right) \ell}}}
\le O_r\pbra{t^{\frac12+\frac1{4r'}}\cdot n^{\frac12-\frac1{4r'}}}^\ell
$$
for low levels of $\ell\le O(r^2)$.
Putting the bound
$$
c(r,r')=\frac{r-r'}{rr'+r/2}
$$
gives the desired bound.

For the special case $r'=1$, we apply the improved Fourier growth bound from \Cref{rmk:tightness}:
$$
L_{1,\ell}(f)
\le O_\ell\pbra{t^{\ell/4}\cdot n^{\ell/4}}
\le O_r\pbra{t^{1/4}\cdot n^{1/4}}^\ell
$$
for low levels of $\ell\le O(r^2)$.
Now \Cref{eq:reduction_to_bs} holds with 
$$
c(r,r')=1-\frac1r
$$
as desired.

\section{Tightness of the Fourier Growth for the Non-Adaptive Case}\label{sec:tightness}

Recall the definition of $k$-fold Forrelation problem from \Cref{def:kforr}.
Here we show the tightness of our Fourier growth bounds for the non-adaptive parallel query algorithms using two-fold Forrelation function:
$$
\Forr_2(x_1,x_2)=\frac1n\sum_{i,j\in[n]}x_1(i)\cdot H_{i,j}\cdot x_2(j).
$$
Since $H$ is the orthonormal Hadamard matrix, each $H_{i,j}$ is $\pm1/\sqrt n$.
As a result, $\Forr_2$ is a degree-$2$ homogeneous function with 
\begin{equation}\label{eq:level-2_forr}
L_{1,2}(\Forr_2)=\sqrt n.
\end{equation}
Let $Q$ be the quantum query algorithm from \Cref{fct:kforr_alg} for $k=2$.
Then its acceptance probability function $f(x)=\Pr\sbra{Q\text{ accepts }x}$ equals $(1+\Forr_2(x))/2$.

Now, for a fixed positive odd number $s$, let $\Maj_s\colon\binpm^s\to\binpm$ be the majority function on $s$ bits.
Define $\Forr_2\circ\Maj_s\colon\binpm^{2sn}\to\Rbb$ by replacing input bits of $\Forr_2$ with majorities on disjoint sets of $s$ bits:
\begin{equation}\label{eq:forrmaj_tightness}
\Forr_2\circ\Maj_s(y)=\frac1n\sum_{i,j\in[n]}\Maj_s(y_{1,i})\cdot H_{i,j}\cdot\Maj_s(y_{2,j}),
\end{equation}
where $y=(y_{1,1},\ldots,y_{1,n},y_{2,1},\ldots,y_{2,n})$ and each $y_{1,i},y_{2,j}\in\binpm^s$.

We substitute the quantum query of $Q$ on $x$ by $s$ parallel queries on $y$.
This produces a non-adaptive quantum query algorithm $\bar Q$ with $s$ parallel queries, and its acceptance probability function is
$$
\bar f(y)=\frac{1+\Forr_2\circ\Maj_s(y)}2.
$$

Now for a fixed positive integer $L$, consider executing $\bar Q$ in parallel on $L$ disjoint inputs and taking the parity of the results.
This is a non-adaptive quantum query algorithm $Q'$ with $t=sL$ parallel queries, and its acceptance probability function is
\begin{equation}\label{eq:f'_tightness}
f'(z)=\frac{1}{2} + \frac{1}{2}\prod_{k\in[L]}\Forr_2\circ\Maj_s(y^k),
\end{equation}
where $z=(y^1,\ldots,y^L)$ and each $y^k=(y_{1,1}^k,\ldots,y_{1,n}^k,y_{2,1}^k,\ldots,y_{2,n}^k)\in\binpm^{2sn}$.

We lower bound the level-$\ell$ Fourier weight of $f'$ with $\ell=2L$.
To this end, we observe that $\hat{\Maj_s}(\emptyset)=0$ and recall that $\Forr_2$ is degree-2 homogeneous.
Since $f'$ is essentially the product of $L$ disjoint copies of $\Forr_2\circ\Maj_s$ (see \Cref{eq:f'_tightness}) and each $\Forr_2\circ\Maj_s$ is a sum of products of two disjoint $\Maj_s$ (see \Cref{eq:forrmaj_tightness}), the level-$\ell$ Fourier coefficients of $f'$ comes from expanding the products of level-$1$ Fourier weight of $\Maj_s$, weighed by the level-$2$ Fourier coefficients of $\Forr_2$.
Therefore
\begin{align*}
L_{1,\ell}(f')
&=\frac12\cdot\pbra{L_{1,2}(\Forr_2)\cdot L_{1,1}(\Maj_s)^2}^L\\
&=\Omega\pbra{\sqrt n\cdot s}^L
\tag{by \Cref{eq:level-2_forr} and $L_{1,1}(\Maj_s)=\Theta(\sqrt s)$}
\end{align*}
Recall that $\ell=2L$, $t=sL$, and $f'$ is a function on $n=2sLn$ input bits.
This implies
$$
L_{1,\ell}(f')\ge\Omega_\ell\pbra{n^{\ell/4}\cdot t^{\ell/4}},
$$
matching the bound in \Cref{rmk:tightness}.

\section{Quantum Query Algorithms with Classical Preprocessing}\label{app:classical_preproc}

In this section, we show that our Fourier analytic approach can be generalized to handle a more general setting, where the quantum query algorithm is allowed to first perform many classical queries as a preprocessing phase.
More precisely, we prove \Cref{lem:classical_preproc} analogous to \Cref{thm:fourier_growth}.
\begin{lemma}\label{lem:classical_preproc}
Let $\Acal$ be an algorithm on $n$-bit inputs:
\begin{itemize}
\item \textsc{Classical Preprocessing Phase.} 
First $\Acal$ performs at most $d$ classical queries.
\item \textsc{Quantum Parallel Query Phase.} 
Then based on the results of the previous phase, $\Acal$ executes a quantum query algorithm $\Bcal$ with arbitrarily many auxiliary qubits and $r$ adaptive rounds of $t\le n$ parallel quantum queries per round.
\end{itemize}
Define $f\colon\binpm^n\to[0,1]$ by $f(x)=\Pr\sbra{\Acal\text{ accepts x}}$.
Then
$$
L_{1,\ell}(f)\le O_{r,\ell}\pbra{(d\cdot t)^\ell\cdot \left(\sqrt{n/t}\right)^{\floorbra{ \left(1-\frac{1}{2r}\right) \ell }}}.
$$
Moreover, this bound holds when some bits of $x$ are fixed in advance.
\end{lemma}

By a similar calculation as in \Cref{app:thm:separation}, \Cref{lem:classical_preproc} strengthens \Cref{thm:separation} as the following theorem.
\begin{theorem}\label{thm:separation_preproc}
For any constant $r\ge2$, the $2r$-fold Forrelation problem on $n$-bit inputs 
\begin{enumerate}
\item can be solved with advantage $2^{-10r}$ by $r$ adaptive rounds of  queries with one quantum query per round, yet
\item any algorithm with $n^{1/(2r)}$ classical preprocessing queries and $r-1$ adaptive quantum query rounds requires $\tilde{\Omega}(n^{1/r^2})$ parallel quantum queries to approximate it.
\end{enumerate}
\end{theorem}

Now we prove \Cref{lem:classical_preproc} which is a simple black-box reduction to \Cref{thm:fourier_growth}.
\begin{proof}[Proof of \Cref{lem:classical_preproc}]
We view the classical preprocessing phase of $\Acal$ as a decision tree $\Dcal$ of depth at most $d$, where each leaf $z$ of $\Dcal$ selects a quantum query algorithm $\Bcal_z$.
In addition, we identify each $z$ as a partial assignment in $\{\pm1,\star\}^n$ where the $\pm1$ values correspond to classical queries and their outcome, and $\star$'s correspond to bits that are not queried in this phase.
In particular, there are at most $d$ non-$\star$ values for each $z$ and we use $z^{-1}(\star)\subseteq[n]$ to denote the entries of these non-$\star$ values.

For each $z$, define $g_z(x)=\Pr[\Bcal_z\text{ accepts }x|z]$ as the acceptance probability function of $\Bcal_z$ conditioned that $x$ is consistent with $z$ on entries in $z^{-1}(\star)$.
By \Cref{thm:fourier_growth}, we have
\begin{equation}\label{eq:lem:classical_preproc_1}
L_{1,k}(g_z)\le O_{r,k}\pbra{t^k\cdot \left(\sqrt{n/t}\right)^{\floorbra{ \left(1-\frac{1}{2r}\right) k }}}
\quad\text{for each $k\ge0$}
\end{equation}
For each $S\subseteq[n]$ of size $\ell$, define $a_S=\sgn(\hat f(S))$ as the sign of the Fourier coefficients at level $\ell$ and use $x_S$ to denote $\prod_{i\in S}x_i$.
Then we have
\begin{align*}
L_{1,\ell}(f)
&=\E_x\sbra{f(x)\sum_{|S|=\ell}a_S\cdot x_S}
=\E_z\sbra{\E_x\sbra{f(x)\sum_{|S|=\ell}a_S\cdot x_S\mid z}}
=\E_z\sbra{\E_x\sbra{g_z(x)\sum_{|S|=\ell}a_S\cdot x_S\mid z}}
\tag{$z$ is sampled by a random root-to-leaf path in $\Dcal$}\\
&=\E_z\sbra{\E_x\sbra{g_z(x)\sum_{\substack{T_1\subseteq z^{-1}(\star)\\T_2\subseteq[n]\setminus z^{-1}(\star)\\|T_1|+|T_2|=\ell}}a_{T_1\cup T_2}\cdot z_{T_1}\cdot x_{T_2}\mid z}}
\tag{by the definition of $z$}\\
&=\E_z\sbra{\sum_{T_1\subseteq z^{-1}(\star),|T_1|\le\ell}z_{T_1}\cdot
\E_x\sbra{g_z(x)\sum_{\substack{T_2\subseteq[n]\setminus z^{-1}(\star)\\|T_2|=\ell-|T_1|}}a_{T_1\cup T_2}\cdot x_{T_2}}}
\tag{by the definition of $g_z$}\\
&\le\E_z\sbra{\sum_{T_1\subseteq z^{-1}(\star),|T_1|\le\ell}L_{1,\ell-|T_1|}(g_z)}
\tag{by the definition of $L_{1,\ell-k}$ and since $z_{T_1}\in\binpm$}\\
&\le\sum_{k=0}^\ell d^k\cdot\E_z\sbra{L_{1,\ell-k}(g_z)}
\tag{since $|z^{-1}(\star)|\le d$}\\
&\le\sum_{k=0}^\ell d^k\cdot O_{r,k}\pbra{t^{\ell-k}\cdot \left(\sqrt{n/t}\right)^{\floorbra{ \left(1-\frac{1}{2r}\right)(\ell-k) }}}
\tag{by \Cref{eq:lem:classical_preproc_1}}\\
&=O_{r,\ell}\pbra{(d\cdot t)^\ell\cdot \left(\sqrt{n/t}\right)^{\floorbra{ \left(1-\frac{1}{2r}\right)\ell}}}
\tag*{\qedhere}
\end{align*}
\end{proof}

\end{document}